\documentclass[preprint,nofootinbib,showkeys]{revtex4-1}

\usepackage{hyperref}
\hypersetup{
  breaklinks=true,
  colorlinks=true,
  linkcolor=blue,
  urlcolor=cyan,
}

\usepackage{physics,amsmath,amssymb,braket,dsfont}
\renewcommand{\t}{\text} 
\newcommand{\f}{\dfrac} 
\newcommand{\p}[1]{\left(#1\right)} 
\renewcommand{\sp}[1]{\left[#1\right]} 
\renewcommand{\set}[1]{\left\{#1\right\}} 
\newcommand{\bk}{\braket} 

\renewcommand{\d}{\text{d}}
\newcommand{\g}{\text{g}}
\newcommand{\e}{\text{e}}
\newcommand{\x}{\text{x}}
\newcommand{\y}{\text{y}}
\newcommand{\z}{\text{z}}

\renewcommand{\c}{\hat{c}}
\newcommand{\n}{\hat{n}}

\newcommand{\A}{\mathcal{A}}
\newcommand{\B}{\mathcal{B}}
\newcommand{\D}{\mathcal{D}}
\newcommand{\E}{\mathcal{E}}
\newcommand{\G}{\mathcal{G}}
\renewcommand{\H}{\mathcal{H}}
\newcommand{\I}{\mathcal{I}}
\newcommand{\K}{\mathcal{K}}
\renewcommand{\L}{\mathcal{L}}

\newcommand{\N}{\mathcal{N}}
\renewcommand{\O}{\mathcal{O}}
\renewcommand{\P}{\mathcal{P}}
\newcommand{\Q}{\mathcal{Q}}
\renewcommand{\S}{\mathcal{S}}
\newcommand{\U}{\mathcal{U}}

\newcommand{\1}{\mathds{1}}

\newcommand{\mA}{m_{\text{A}}} 

\usepackage{tikz}
\newcommand{\lvec}[1]
{\reflectbox{\ensuremath{\vec{\reflectbox{\ensuremath{#1}}}}}}

\usepackage{graphicx} 
\usepackage{grffile} 
\graphicspath{{./figures/}} 
\usepackage[caption=false]{subfig} 

\usepackage[inline]{enumitem}
\setlist[enumerate,1]{label={(\roman*)}}

\usepackage{tikz,tikz-feynman}
\tikzset{
  baseline = (current bounding box.center)
}
\tikzfeynmanset{
  compat = 1.1.0,
  every feynman = {/tikzfeynman/small}
}
\newcommand{\shrink}[1]{\scalebox{0.8}{#1}} 

\usepackage{xcolor}
\definecolor{lightblue}{RGB}{31,119,180}
\definecolor{orange}{RGB}{255,127,14}
\definecolor{green}{RGB}{44,160,44}
\definecolor{lightred}{RGB}{214,39,40}

\makeatletter
\def\mathcolor#1#{\@mathcolor{#1}}
\def\@mathcolor#1#2#3{
  \protect\leavevmode
  \begingroup
    \color#1{#2}#3
  \endgroup
}
\makeatother
\newcommand{\bmu}{\mathcolor{lightblue}{\mu}}
\newcommand{\onu}{\mathcolor{orange}{\nu}}
\newcommand{\grho}{\mathcolor{green}{\rho}}
\newcommand{\re}{\mathcolor{lightred}{\text{e}}}

\usepackage[normalem]{ulem}

\begin{document}

\title{Effective multi-body SU($N$)-symmetric interactions of
  ultracold fermionic atoms on a 3-D lattice}

\author{M A Perlin}
\email{mika.perlin@gmail.com}
\author{A M Rey}
\affiliation{JILA, National Institute of Standards and Technology and
  University of Colorado, 440 UCB, Boulder, Colorado 80309, USA}
\affiliation{Center for Theory of Quantum Matter, 440 UCB, Boulder,
  Colorado 80309, USA}
\affiliation{Department of Physics, University of Colorado, 390 UCB,
  Boulder, Colorado 80309, USA}

\begin{abstract}
  Rapid advancements in the experimental capabilities with ultracold
  alkaline-earth-like atoms (AEAs) bring to a surprisingly near term
  the prospect of performing quantum simulations of spin models and
  lattice field theories exhibiting SU($N$) symmetry.  Motivated in
  particular by recent experiments preparing high density samples of
  strongly interacting ${}^{87}$Sr atoms in a three-dimensional
  optical lattice, we develop a low-energy effective theory of
  fermionic AEAs which exhibits emergent multi-body SU($N$)-symmetric
  interactions, where $N$ is the number of atomic nuclear spin levels.
  Our theory is limited to the experimental regime of
  \begin{enumerate*}
  \item a deep lattice, with
  \item at most one atom occupying each nuclear spin state on any
    lattice site.
  \end{enumerate*}
  The latter restriction is a consequence of initial ground-state
  preparation.  We fully characterize the low-lying excitations in our
  effective theory, and compare predictions of many-body interaction
  energies with direct measurements of many-body excitation spectra in
  an optical lattice clock.  Our work makes the first step in enabling
  a controlled, bottom-up experimental investigation of multi-body
  SU($N$) physics.
\end{abstract}


\maketitle

\tableofcontents

\section{Introduction}
\label{sec:introduction}

Fermionic alkaline-earth atoms (AEAs), in addition to other atoms such
as ytterbium (Yb) sharing similar electronic structure, are currently
the building blocks of the most precise atomic clocks in the world
\cite{derevianko2011colloquium, katori2011optical, ludlow2015optical}.
These atoms have a unique, ultra-narrow optical transition between
metastable ${}^1S_0$ and ${}^3P_0$ electronic orbital states, i.e.~the
``clock states'', that allows for coherence times which can exceed 100
seconds \cite{porsev2004hyperfine, santra2004properties}.
Furthermore, AEAs can be trapped in fully controllable optical lattice
potentials and interrogated with ultra-stable lasers that can resolve
and probe their rich hyperfine spectra, consisting of $N$ different
nuclear spin levels with $N$ as large as 10 in strontium (${}^{87}$Sr)
and 6 in ytterbium (${}^{173}$Yb).

In 2015 the ${}^{87}$Sr optical lattice clock (OLC) at JILA, operated
in a one-dimensional (1-D) lattice at microkelvin temperatures,
achieved a total fractional uncertainty of $2\times10^{-18}$
\cite{bloom2014optical, nicholson2015systematic}.  More recently
(2017), a new generation of OLCs became operational at JILA,
interrogating a Fermi degenerate gas of ${}^{87}$Sr atoms in a 3-D
lattice at nanokelvin temperatures \cite{campbell2017fermidegenerate}.
All of these atoms' degrees of freedom, including the electronic
orbital, nuclear spin, and motional states, can be fully controlled
with high fidelity in a 3-D lattice \cite{daley2008quantum,
  gorshkov2009alkalineearthmetal, daley2011quantum,
  daley2011statedependent}.  With frequency measurements reaching the
$10^{-19}$ fractional uncertainty level, the new OLCs are thus
enabling an exciting opportunity to probe, for the first time, quantum
dynamics with sub-millihertz spectral resolution
\cite{campbell2017fermidegenerate}.

A wonderful consequence of the efforts to build better clocks is the
development of highly controllable quantum simulators of many-body
systems in the strongly-interacting regime, where inter-particle
interactions set the largest energy scale relevant for system dynamics
\cite{swallows2011suppression, lemke2011wave,
  campbell2017fermidegenerate}.  The marriage between precision clock
spectroscopy and quantum many-body physics \cite{taie2012su,
  martin2013quantum, scazza2014observation, cappellini2014direct,
  zhang2014spectroscopic, rey2014probing} has an enormous potential to
enable novel explorations of physics for the same reason that makes
AEAs such remarkable time-keepers.  Specifically, due to the lack of
electronic orbital angular momentum in the ${}^1S_0$ and ${}^3P_0$
states, AEAs exhibit decoupled orbital and nuclear spin degrees of
freedom.  For atoms with $N$ nuclear spin levels, this decoupling
leads to nuclear-spin-conserving SU($N$)-symmetric interactions
governed entirely by orbital-state parameters
\cite{cazalilla2009ultracold, taie2012su, zhang2014spectroscopic}.

The presence of this exotic SU($N$) symmetry in a highly controllable
experimental platform opens the door to experimental studies of
e.g.~the SU($N$) Heisenberg model, whose phase diagram is believed to
exhibit features such as a chiral spin liquid (CSL) phase with
topological order and fractional statistics \cite{hermele2009mott,
  hermele2011topological, chen2016syntheticgaugefield}.  In addition
to illuminating open questions in our understanding of the fractional
quantum Hall effect and unconventional superconductivity
\cite{lee2006doping, lee2008high, gong2014emergent}, the CSL can
support non-Abelian excitations which allow for universal topological
quantum computation \cite{freedman2004class, hermele2011topological}.
Harnessing the SU($N$)-symmetric interactions of AEAs might also
enable the simulation of various lattice gauge theories
\cite{wieseu.-j.2013ultracold, zohar2016quantum}, some of which share
important qualitative features with quantum chromodynamics such as
few-body bound states and confinement \cite{banerjee2013atomic,
  rico2018nuclear}.  These direct, quantum simulations have an
extraordinary potential to provide novel insights by circumventing
e.g.~severe sign problems which plague classical simulations of
strongly interacting fermionic systems \cite{wieseu.-j.2013ultracold,
  wu2003exact}.

In this work, we investigate the first experimental capabilities with
ultracold fermionic AEAs to prepare high-density samples in a 3-D
lattice with multiple occupation of individual lattice sites
\cite{goban2018emergence}.  Specifically, we consider ground-state
preparation of isolated few-body systems in the deep-lattice limit,
and carry out a bottom-up investigation of emergent multi-body
interactions on multiply-occupied lattice sites.  These multi-body
interactions appear in a low-energy effective theory of the atoms, and
inherit the SU($N$) symmetry of their bare, pair-wise interactions,
thereby enabling experimental studies of multi-body SU($N$) physics
through the exquisite capabilities with OLCs.  Our theory is limited
to the experimental regime of at most one atom occupying each nuclear
spin state on any lattice site, which is a consequence of the
experimental protocol which starts with all atoms in the ground state.

Though effective multi-body interactions have previously been studied
in the context of harmonically \cite{johnson2012effective,
  yin2014universal} and lattice-confined \cite{johnson2009effective}
neutral bosons prepared in a single hyperfine state, our work deals
for the first time with fermions that have internal degrees of freedom
and multiple collisional parameters.  Some past work has detected
experimental signatures of multi-body interactions in the form of
quantum phase revivals \cite{will2010timeresolved}.  We instead
compare the many-body interaction energies predicted by our low-energy
effective theory to the experimental measurements of the
density-dependent orbital excitation spectra performed in
ref.~\cite{goban2018emergence}, similarly to the measurements with
bosons performed in ref.~\cite{mark2011precision}.  To facilitate this
comparison of excitation spectra and to characterize the low-lying
excitations in our effective theory, we consider a restriction of our
theory to states with at most one orbital excitation per lattice site.
In this case, we find that the SU($N$) symmetry of atomic collisions
allow the effective multi-body interactions to take a remarkably
simple form.

The remainder of this paper is structured as follows.  In section
\ref{sec:overview} we summarize the experimental procedures relevant
to our work, provide an overview of the one- and two-body physics of
ultracold atoms in a deep lattice, and preview our main technical
results.  In section \ref{sec:low_energy} we discuss our method for
deriving a low-energy effective theory, provide a perturbative
expansion for the net effective Hamiltonian, and compute all $M$-body
Hamiltonians through third order in the low-energy effective theory.
We then analyze the low-lying excitations of the effective theory in
section \ref{sec:spectra}, comparing spectral predictions with
experimental measurements, and study the orbital-state dynamics of
nuclear spin mixtures interrogated via Rabi spectroscopy.  Finally, we
summarize and conclude our findings in section \ref{sec:summary}, and
provide some discussion of future outlooks.

\section{Background and overview}
\label{sec:overview}

The work in this paper is closely tied to the experimental work
reported in ref.~\cite{goban2018emergence}; we begin with a short
summary of the relevant experimental procedures therein.  The
experiment begins by preparing a degenerate gas of $10^4$-$10^5$
(fermionic) ${}^{87}$Sr atoms in a uniform mixture of their ten
nuclear spin states and at $\sim0.1$ of their fermi temperature
($\sim10$ nanokelvin) \cite{campbell2017fermidegenerate,
  marti2018imaging}.  This gas is loaded into a primitive cubic
optical lattice at the ``magic wavelength'' for which both ground
(${}^1S_0$) and first-excited (${}^3P_0$) electronic orbital states of
the atoms experience the same lattice potential \cite{ye2008quantum}.
Lattice depths along the principal axes of the lattice are roughly
equal in magnitude, with a geometric mean that can be varied from 30
to 80 $E_{\t{R}}$, where $E_{\t{R}}\approx3.5\times2\pi~\t{kHz}$ is
the lattice photon recoil energy of the atoms (with the reduced Planck
constant $\hbar=1$ throughout this paper).  These lattice depths are
sufficiently large as to neglect tunneling on the time scales relevant
to the experiment.  The temperature of the atoms is also low enough to
neglect thermal occupation of motional states outside the ground-state
manifold.

Once loaded into an optical lattice, atoms are addressed by an
external (``clock'') interrogation laser with an ultranarrow (26 mHz)
linewidth, detuned by $\Delta$ from the single-atom ${}^1S_0-{}^3P_0$
transition frequency $\omega_0$.  After a fixed interrogation time,
the experiment turns off the interrogation laser, removes all
ground-state (${}^1S_0$) atoms from the lattice, and uses absorption
imaging to count the remaining excited-state (${}^3P_0$) atoms.
Non-interacting atoms in singly-occupied lattice sites feature the
typical single particle lineshape peaked at $\Delta=0$.  The
lineshapes of multiply-occupied lattice sites, meanwhile, are shifted
by inter-atomic interactions, which results in spectroscopic peaks
(i.e.~local maxima in excited-state atom counts) away from $\Delta=0$.
A sweep across different detunings $\Delta$ (on the scale of
inter-atomic interaction energies) thus constitutes a measurement of
the many-body orbital excitation spectrum.  We note that this
spectroscopic protocol addresses only singly-excited orbital states of
lattice sites.  Doubly-excited states are off resonant due to
\begin{enumerate*}
\item the interaction-induced non-linearity ($\sim$kHz) of the orbital
  excitation energies, and
\item the ultranarrow linewidth ($\sim$mHz) of the interrogation
  laser.
\end{enumerate*}

Although an external trapping potential will generally break discrete
translational symmetry of the lattice, any background inhomogeneity
can be made negligible by spectroscopically addressing a sufficiently
small region of the lattice \cite{goban2018emergence}.  Throughout
this paper, we work strictly in the deep-lattice regime with
negligible tunneling between lattice sites.  We also neglect any
lattice inhomogeneities and assume that both atomic orbital states
(i.e.~${}^1S_0$ and ${}^3P_0$) experience identical lattice
potentials.  The single-particle Hamiltonian of the atoms can then be
written in the form
\begin{align}
  H_0 = \sum_{i,n,\mu,s} E_n \c_{in\mu s}^\dag \c_{in\mu s},
  \label{eq:H_0}
\end{align}
where $\c_{in\mu s}$ is a fermionic operator which annihilates a
single atom on lattice site $i\in\mathbb{Z}^3$ in motional state
$n\in\mathbb{N}_0^3$ with nuclear spin $\mu\in\set{-I,-I+1,\cdots,I}$
(i.e.~projected onto a quantization axis) and orbital state
$s\in\set{\g,\e}$; and $E_n$ is the energy of a single atom in
motional state $n$.  In a harmonic trap approximation we would have
$E_n=\p{3/2+n_\x+n_\y+n_\z}\omega$ for an on-site angular trap
frequency $\omega$, but in general the aharmonicity of the lattice
potential will cause a non-negligible shift in motional state
energies.

In the absence of hyperfine coupling, as when addressing the spinless
${}^1S_0$ ($\g$) and ${}^3P_0$ ($\e$) orbital states of AEAs,
interactions between any two atoms are governed by their orbital
states alone, and are therefore characterized by four scattering
lengths $a_X$ with $X\in\set{\g\g,\e\g^-,\e\g^+,\e\e}$, where the $+$
($-$) superscript denotes symmetrization (anti-symmetrization) of a
two-body orbital state under particle exchange.  In the low-energy
limit, we can write the bare two-body interaction Hamiltonian in the
form \cite{gorshkov2010twoorbital}
\begin{align}
  H_{\t{int}} = \sum_{\substack{\mu<\nu\\s}}
  G_s \int \d^3x~ \hat\rho_{\mu s} \hat\rho_{\nu s}
  + G_+ \sum_{\mu,\nu}
  \int \d^3x~ \hat\rho_{\mu,\e} \hat\rho_{\nu,\g}
  + G_- \sum_{\mu,\nu} \int \d^3x~
  \hat\psi_{\mu,\e}^\dag \hat\psi_{\nu,\g}^\dag
  \hat\psi_{\nu,\e} \hat\psi_{\mu,\g},
  \label{eq:H_int_start}
\end{align}
where $\hat\psi_{\mu s}$ is a fermionic field operator for atoms with
nuclear spin $\mu$ and orbital state $s$;
$\hat\rho_{\mu s}\equiv\hat\psi_{\mu s}^\dag\hat\psi_{\mu s}$ is an
atomic density field operator; and the coupling constants $G_X$ are
defined in terms of the scattering lengths $a_Y$ by
\begin{align}
  G_{s=\g,\e} \equiv \f{4\pi}{\mA}~ a_{ss}, &&
  G_\pm \equiv \f{2\pi}{\mA}~ \p{a_{\e\g+} \pm a_{\e\g-}},
  \label{eq:couplings}
\end{align}
where $\mA$ is the mass of a single atom.  Defining for brevity
\begin{align}
  G^{qr}_{st} \equiv \left\{
    \begin{array}{ll}
      G_q & ~ q = r = s = t \\
      G_+ & ~ q \ne r ~ \t{and} ~ (q,r) = (s,t) \\
      G_- & ~ q \ne r ~ \t{and} ~ (q,r) = (t,s) \\
      0 & ~ \t{otherwise}
    \end{array}\right.,
  \label{eq:coupling_tensor}
\end{align}
where $q,r,s,t\in\set{\g,\e}$ are orbital state indices, we can
alternately write the bare two-body interaction Hamiltonian in the
more compact form
\begin{align}
  H_{\t{int}}
  = \f12 \sum_{\substack{q,r,s,t\\\mu,\nu}} G^{qr}_{st} \int \d^3x~
  \hat\psi_{\mu s}^\dag \hat\psi_{\nu t}^\dag
  \hat\psi_{\nu r} \hat\psi_{\mu q}.
  \label{eq:H_int_fields}
\end{align}
For nuclear spins $\mu,\nu$, the symbol $G^{qr}_{st}$ gives the
coupling constant between the two-atom states
$(\mu,q)+(\nu,r)\leftrightarrow(\mu,s)+(\nu,t)$.

Note that the Hamiltonian in \eqref{eq:H_int_fields} is not the true
microscopic interaction Hamiltonian of AEAs, but rather a generic form
for a low-energy effective field theoretic description of two-body
interactions \cite{scazza2014observation, cazalilla2009ultracold,
  johnson2012effective, yin2014universal, johnson2009effective,
  gorshkov2010twoorbital, wall2013strongly, busch1998two}.  There are
therefore two important points to keep in mind concerning our use of
\eqref{eq:H_int_fields} to describe two-body interactions.  First, the
use of effective field theory generically gives rise to divergences
that must be dealt with either through regularization, e.g.~of the
zero-range interaction potential implicitly assumed in the expression
of \eqref{eq:H_int_fields} \cite{giorgini2008theory}, or through
renormalization of the coupling constants in the theory.  We chose the
latter approach, as we will in any case find it convenient to
renormalize the coupling constants in the effective theory developed
in section \ref{sec:low_energy}.  The choice of method to regulate
divergences has no effect on the underlying physics.

Second, \eqref{eq:H_int_fields} is only the first term in a low-energy
expansion of two-body interactions in effective field theory, which
generally includes additional terms containing derivatives of field
operators.  Derivative terms correspond to the dependence of two-body
scattering on the relative momentum $k$ of particles involved, with
$k\to0$ in the zero-energy limit.  In the present case of $s$-wave
scattering, the leading dependence of the two-body interaction
Hamiltonian on the relative momentum $k$ can be captured by use of an
energy-dependent pseudo-potential, which amounts to using a
$k$-dependent effective scattering length \cite{blume2002fermi}.  This
effective scattering length can be determined by expanding the
$s$-wave collisional phase shift in powers of the relative momentum
$k$ \cite{giorgini2008theory, flambaum1999analytical}.  Details of
this expansion will depend on the characteristic length scale of
finite-range interactions.  In our work, these corrections to
\eqref{eq:H_int_fields} will be relevant only for the calculation of
two-body interaction energies, appearing at third order in the
coupling constants $G_X$.  As we are primarily interested in $M$-body
interactions for $M\ge3$, we defer this calculation to Appendix
\ref{sec:momentum_dependence}.  We note that our approach of using an
unregularized contact potential, renormalizing coupling constants, and
separately accounting for momentum-dependent scattering is essentially
the same as the approach used for similar calculations in
refs.~\cite{johnson2012effective, yin2014universal,
  johnson2009effective}.  While this approach does not provide insight
into the microscopic structure of inter-atomic interactions, it is
suitable for the phenomenological description of these interactions,
and particular for our eventual development of a low-energy effective
theory.

We now expand the field operators $\hat\psi_{\mu s}$ in the Wannier
basis for a 3-D lattice, such that
$\hat\psi_{\mu s}(x) = \sum_{i,n} \phi_{in}(x) \c_{in\mu s}$ with
spatial wavefunctions $\phi_{in}$ and fermionic annihilation operators
$\c_{in\mu s}$ indexed by lattice sites $i$ and motional states $n$.
Invoking the tight-binding approximation, we assume that the spatial
overlap integral in \eqref{eq:H_int_fields} is negligible unless all
wavefunctions are localized at the same lattice site; we discuss the
breakdown of this approximation and its consequences for our
low-energy effective theory in Appendix \ref{sec:error}.  The relevant
spatial overlap integral is then
\begin{align}
  K^{k\ell}_{mn}
  \equiv \int \d^3x~ \phi_{im}^* \phi_{in}^* \phi_{i\ell} \phi_{ik},
  \label{eq:K_klmn}
\end{align}
which for a lattice with discrete translational invariance is
independent of the lattice site $i$.  The two-body interaction
Hamiltonian can be written in terms of this overlap integral as
\begin{align}
  H_{\t{int}}
  = \f12 \sum_{\substack{i,k,\ell,m,n\\q,r,s,t\\\mu,\nu}}
  K^{k\ell}_{mn} G^{qr}_{st}
  \c_{im\mu s}^\dag \c_{in\nu t}^\dag \c_{i\ell\nu r} \c_{ik\mu q}
  \equiv \f12 \sum K^{k\ell}_{mn} G^{qr}_{st}
  \c_{m\mu s}^\dag \c_{n\nu t}^\dag \c_{\ell\nu r} \c_{k\mu q},
  \label{eq:H_int}
\end{align}
where for brevity we will henceforth suppress the identical site index
($i$) on all operators, and implicitly sum over all free indices in a
summand (i.e.~indices which do not have a fixed value).  We may also
at times suppress motional state indices on the overlap integral
$K^{k\ell}_{mn}$, in which case the suppressed indices are implicitly
zero (corresponding to a motional ground state); i.e.
\begin{align}
  K^{\ell m}_n \equiv K^{\ell m}_{n,0},
  &&
  K^m_n \equiv K^{m,0}_{n,0},
  &&
  K_{mn} \equiv K^{0,0}_{mn},
  &&
  K_n \equiv K^{0,0}_{n,0},
  &&
  K \equiv K^{0,0}_{0,0}.
  \label{eq:K}
\end{align}
For simplicity, we will also generally work in a gauge for which all
two-body overlap integrals are real, such that
$K^{k\ell}_{mn}={K^{k\ell}_{mn}}^*=K^{mn}_{k\ell}$.  The existence of
such a gauge is guaranteed by the analytic properties of the Wannier
wavefunctions $\phi_{in}$ \cite{kohn1959analytic}.

Current experiments with ${}^{87}$Sr can prepare up to five atoms in
the same (ground) orbital state on a single lattice site, and
coherently address states with a single orbital excitation per lattice
site \cite{goban2018emergence}.  At ultracold temperatures well below
the non-interacting motional excitation energies
$\Delta_n\equiv E_n-E_0$ for $n>0$, atoms only occupy their motional
ground state in the lattice.  For this reason, it is common to map the
description of these atoms onto a single-band Hubbard model that
captures all dynamics within the subspace of motional ground states of
non-interacting atoms, i.e.~with wavefunctions $\phi_{i,0}$.
Interactions, however, modify atoms' motional ground-state
wavefuctions.  The true motional ground state of a collection of
interacting atoms is then an admixture of the non-interacting motional
eigenstates, and a naive Hubbard model that assumes atomic
wavefunctions $\phi_{i,0}$ will fail to reproduce the interacting
atoms' orbital excitation spectrum.  Formally, corrections to the
spectrum of interacting atoms can be accounted for by a perturbative
treatment of far-off-resonant terms in the interaction Hamiltonian of
\eqref{eq:H_int} that create atoms in excited motional states,
e.g.~$\sim\c_{n\mu s}^\dag \c_{0,\nu t}^\dag \c_{0,\nu r} \c_{0,\mu
  q}$ with $n>0$.  These corrections can be understood through
interaction-induced {\it virtual} occupation of higher bands
(i.e.~excited motional states), which becomes relevant as more atoms
occupy the same lattice site, such that their interaction energy
becomes non-negligible compared to the motional excitation energies
$\Delta_n$.

In order to recast interaction-induced modifications to orbital
excitation spectra as corrections to the simple Hubbard model
(i.e.~computed using the non-interacting ground-state wavefunctions
$\phi_{i,0}$), we develop a low-energy effective theory of interacting
AEAs in a deep lattice.  To simplify our theory, we assume that any
$N$ atoms on a single lattice site occupy distinct nuclear spin
states.  This assumption applies for any experimental protocol in
which all atoms are initially prepared in their orbital and motional
ground states (as e.g.~in ref.~\cite{goban2018emergence}).  In this
case, multiple occupation of a single nuclear spin state on any given
lattice site is initially forbidden by fermionic statistics.
Subsequent violation of this condition cannot occur in the absence of
inter-site effects or hyperfine coupling between nuclear spin states,
as is the case of the experiment in ref.~\cite{goban2018emergence}.

Our low-energy effective theory exhibits SU($N$)-symmetric multi-body
interactions, such that the effective interaction Hamiltonian can be
written in the form
\begin{align}
  H_{\t{int}}^{\t{eff}} = \sum_{M=2}^{2I+1} \sum_{p\ge1} H_M^{(p)},
\end{align}
where $H_M^{(p)}$ is an $M$-body Hamiltonian of order $p$ in the
coupling constants $G_X$, and $I$ is the total nuclear spin of each
atom (e.g.~$I=9/2$ for ${}^{87}$Sr).  The sum terminates at $2I+1$
because this is the largest number of atoms which may initially occupy
a single lattice site.  We explicitly compute all $M$-body
Hamiltonians $H_M\equiv\sum_p H_M^{(p)}$ through order $p=3$, yielding
effective two-, three-, and four-body interactions.  To
\begin{enumerate*}
\item facilitate a comparison with the experimental measurements of
  many-body orbital excitation spectra performed in
  ref.~\cite{goban2018emergence} and
\item characterize the low-lying excitations in our effective theory,
\end{enumerate*}
we additionally restrict the multi-body Hamiltonians $H_M$ to states
with at most one orbital excitation per lattice site.  Under this
restriction, we find that the SU($N$) symmetry of atomic collisions
allows us to express all multi-body Hamiltonians in the simple form
\begin{align}
  H_M = \sum_{\abs{\set{\mu_j}}=M}
  H_2^{(\mu_1,\mu_2)} \prod_{\alpha=3}^M \n_{\mu_\alpha,\g},
  \label{eq:H_M_preview}
\end{align}
where $H_2^{(\mu_1,\mu_2)}$ is a two-body Hamiltonian addressing atoms
with nuclear spin $\mu_1,\mu_2$; and
$\n_{\mu s}=\c_{\mu s}^\dag\c_{\mu s}$ is a number operator for atoms
with nuclear state $\mu$ and orbital state $s$.  The sum in
\eqref{eq:H_M_preview} is performed over all choices of nuclear spins
$\mu_j$ with $j=1,2,\cdots,M$ for which all $\mu_j$ are distinct, or
equivalently all choices of $\mu_j$ for which the set $\set{\mu_j}$
contains $M$ elements, for a total of ${2I+1\choose M}\times\p{M!}$
nuclear spin combinations.  The key feature of the $M$-body
interactions in \eqref{eq:H_M_preview} is that they ultimately take
the same form as two-body interactions, but with the addition of $M-2$
spectator atoms.  This form is a direct consequence of the SU($N$)
symmetry of underlying two-body interactions.

\section{Low-energy effective theory}
\label{sec:low_energy}

The net Hamiltonian $H = H_0 + H_{\t{int}}$ for interacting AEAs on a
lattice is not diagonal with respect to single-particle motional state
indices (e.g.~$n\in\mathbb{N}_0^3$).  The problem of determining
interacting atoms' orbital excitation spectrum therefore nominally
involves all atomic motional degrees of freedom.  At zero temperature,
however, each orbital state of a collection of interacting atoms is
associated with a single motional ground state.  In order to compute
an orbital excitation spectrum at zero temperature, in principle we
need to identify this motional ground state.  We can then ignore all
excited motional states, which will be neither thermally occupied nor
externally interrogated.  Such a procedure would drastically reduce
the dimensionality of the Hilbert space necessary to describe the
atoms, thereby greatly simplifying any description of the atoms'
orbital spectrum and internal (i.e.~nuclear and orbital) dynamics.  In
practice, however, identifying the motional ground states of
interacting atoms and writing down a Hamiltonian restricted to this
subspace is a very difficult process to carry out analytically.

We denote the motional ground-state subspace of the non-interacting
Hamiltonian $H_0$ by $\H_{\t{ground}}^{\t{single}}$, and the motional
ground-state subspace of the interacting Hamiltonian
$H = H_0 + H_{\t{int}}$ by $\H_{\t{ground}}^{\t{multi}}$.  That is,
all atomic wavefunctions for states within
$\H_{\t{ground}}^{\t{single}}$ are described by $\phi_{i,0}$, while
the atomic wavefunctions for states within
$\H_{\t{ground}}^{\t{multi}}$ are generally unknown, and are in
principle determined by minimizing the energy of a state with respect
to its motional degrees of freedom.  Both
$\H_{\t{ground}}^{\t{single}}$ and $\H_{\t{ground}}^{\t{multi}}$ are
subspaces of the full Hilbert space $\H_{\t{full}}$.  When
interactions are sufficiently weak compared to the spectral gap
$\Delta$ between $\H_{\t{ground}}^{\t{single}}$ and its orthogonal
complement $\H_{\t{full}}\setminus\H_{\t{ground}}^{\t{single}}$, one
can identify a particular unitary operator $U$ (acting on the full
Hilbert space $\H_{\t{full}}$) which rotates
$\H_{\t{ground}}^{\t{multi}}$ into $\H_{\t{ground}}^{\t{single}}$
\cite{bravyi2011schrieffer}.  This unitary $U$ can be used to
construct an {\it effective Hamiltonian} $H_{\t{eff}} = U H U^\dag$
with two key properties:
\begin{enumerate*}
\item $H_{\t{eff}}$ is diagonal in the same (known) basis as the
  non-interacting Hamiltonian $H_0$, and
\item the spectrum of $H_{\t{eff}}$ on $\H_{\t{ground}}^{\t{single}}$
  is identical to that of the interacting Hamiltonian $H$ on
  $\H_{\t{ground}}^{\t{multi}}$.
\end{enumerate*}
The use of an effective Hamiltonian $H_{\t{eff}}$ thus overcomes the
need to identify $\H_{\t{ground}}^{\t{multi}}$ in order to compute the
orbital spectrum of $H$ at zero temperature.  This method for
constructing an effective theory is commonly known as the
Schrieffer-Wolff transformation, named after the authors of its
celebrated application in relating the Anderson and Kondo models of
magnetic impurities in metals \cite{schrieffer1966relation}.

Using the machinery developed in ref.~\cite{bravyi2011schrieffer} for
performing a rotation between low-energy subspaces of a perturbed
(i.e.~interacting) and unperturbed (i.e.~non-interacting) Hamiltonian,
we derive an expansion for an effective interaction Hamiltonian
$H_{\t{int}}^{\t{eff}}$ in terms two-body interaction Hamiltonian
$H_{\t{int}}$ (see Appendix \ref{sec:eff_derivation}).  This expansion
takes the form
\begin{align}
  H_{\t{int}}^{\t{eff}} = \sum_{p\ge1} H_{\t{int}}^{(p)},
  \label{eq:H_int_eff}
\end{align}
where $H_{\t{int}}^{(p)}$ is order $p$ in $H_{\t{int}}$.  Letting
$\E_0\equiv \H_{\t{full}}\setminus\H_{\t{ground}}^{\t{single}}$ denote
the orthogonal complement of $\H_{\t{ground}}^{\t{single}}$
(i.e.~$\E_0$ is the space of all states with at least one atom in an
excited motional state), $\B_0\p{\E_0}$ denote an eigenbasis of $\E_0$
with respect to the single-particle Hamiltonian $H_0$, and $E_\alpha$
denote the motional energy (with respect to $H_0$) of a state
$\ket\alpha\in\E_0$ relative to the corresponding motional
ground-state energy, we define the operator
\begin{align}
  \I \equiv \sum_{\ket\alpha\in\B_0\p{\E_0}} \f{\op\alpha}{E_\alpha},
\end{align}
which sums over projections onto excited states with corresponding
energetic suppression factors.  The operator $\I$ together with the
projector $\P_0$ onto $\H_{\t{ground}}^{\t{single}}$ allows us
concisely write the first few terms in \eqref{eq:H_int_eff} as
\begin{align}
  H_{\t{int}}^{(1)} = \P_0 H_{\t{int}} \P_0,
  &&
  H_{\t{int}}^{(2)} = -\P_0 H_{\t{int}} \I H_{\t{int}} \P_0,
  \label{eq:H_int_1_2}
\end{align}
\begin{align}
  H_{\t{int}}^{(3)}
  = \P_0 H_{\t{int}} \I H_{\t{int}} \I H_{\t{int}} \P_0
  - \f12\sp{\P_0 H_{\t{int}} \P_0,
    \P_0 H_{\t{int}} \I^2 H_{\t{int}} \P_0}_+,
  \label{eq:H_int_3}
\end{align}
where $\sp{X,Y}_+\equiv XY+YX$.  Writing down a single-band Hubbard
model that simply neglects excited atomic motional states and uses
$H_{\t{int}}$ directly to describe the orbital spectrum of interacting
atoms is thus equivalent to truncating our expansion for
$H_{\t{int}}^{\t{eff}}$ at first order.  In addition to this first
order term, the expansion involves {\it effective corrections} to the
action of $H_{\t{int}}$ on the non-interacting motional ground states
(i.e.~on $\H_{\t{ground}}^{\t{single}}$) in the form of higher-order
terms with {\it intermediate} or {\it virtual occupation} of excited
states, manifest in $\I$.

Substituting the definition of $\I$ into
\eqref{eq:H_int_1_2}-\eqref{eq:H_int_3} yields expressions that are
highly reminiscent of standard non-degenerate perturbation theory in
quantum mechanics, but which nonetheless exhibit crucial differences.
The first, and most obvious difference is that these expressions are
operator equations, and that the sums over virtual states are
performed over a basis for the orthogonal complement of the subspace
$\H_{\t{ground}}^{\t{single}}$, rather than a basis for the orthogonal
complement of a single state, as in non-degenerate perturbation
theory.  Second, the non-degeneracy condition in standard perturbation
theory is here elevated to a restriction on the magnitude of the
perturbation $H_{\t{int}}$ relative to the spectral gap $\Delta$ of
the non-interacting Hamiltonian $H_0$ between
$\H_{\t{ground}}^{\t{single}}$ and the excited subspace $\E_0$.
Specifically, the validity of \eqref{eq:H_int_1_2}-\eqref{eq:H_int_3}
is conditional only on $\norm{H_{\t{int}}}\le\Delta/2$, where
$\norm{X}\equiv\max_{\ket\psi\in\H}\sqrt{\bk{\psi|X^\dag X|\psi}}$ is
the operator norm, with no restrictions on spectral gaps or
degeneracies within $\H_{\t{ground}}^{\t{single}}$ \cite{davis1969new,
  bravyi2011schrieffer}.  Finally, the effective theory involves no
corrections to the non-interacting many-body energy eigenstates; the
purpose of constructing the effective Hamiltonian
$H_{\t{eff}} = H_0 + H_{\t{int}}^{\t{eff}}$ is to reproduce, on the
known eigenstates of the non-interacting Hamiltonian $H_0$ within
$\H_{\t{ground}}^{\t{single}}$, the spectrum of the interacting
Hamiltonian $H = H_0 + H_{\t{int}}$ on $\H_{\t{ground}}^{\t{multi}}$.
``Correcting'' the eigenstates of the non-interacting Hamiltonian
$H_0$ on $\H_{\t{ground}}^{\t{single}}$ thus invalidates the effective
theory.

As a last comment, we note that our chosen method for constructing an
effective Hamiltonian is distinct from adiabatic elimination methods
which are commonly used in the atomic physics and quantum optics
communities to develop effective theories for e.g.~the low-lying
levels of a Lambda system \cite{brion2007adiabatic,
  james2007effective, reiter2012effective, sanz2016adiabatic}.  Unlike
the perturbative, but exact Schieffer-Wolff transformation, adiabatic
elimination methods use approximations which rely on the fast dynamics
of excited states.  While generally reasonable, these approximations
must be made carefully to avoid potential problems with
self-consistency (see section 3 of ref.~\cite{brion2007adiabatic}),
and yield no obvious or straightforward means to compute effective
corrections beyond second order in the couplings between low- and
high-energy sectors of a Hilbert space \cite{james2007effective,
  reiter2012effective, sanz2016adiabatic}.  While at least one attempt
at systematically computing higher-order corrections in the framework
of adiabatic elimination has recently been made
\cite{sanz2016adiabatic}, the resulting expressions do not lend
themselves as nicely to analytical treatment, and were in any case
found by the authors to be equivalent to a Schrieffer-Wolff
transformation.

\subsection{Diagrammatic representation of effective Hamiltonians}

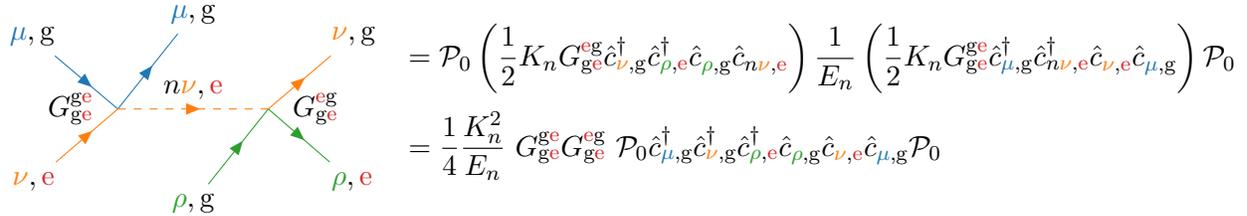
\begin{figure}
  \centering
  \(
  \begin{tikzpicture}
    \begin{feynman}
      \vertex (v1);
      \vertex[above left = of v1] (f1) {$\bmu,\g$};
      \vertex[below left = of v1] (f2) {$\onu,\re$};
      \vertex[right = of v1] (vm);
      \vertex[right = of vm] (v2);
      \vertex[above = of vm] (f3) {$\bmu,\g$};
      \vertex[below = of vm] (f4) {$\grho,\g$};
      \vertex[below right = of v2] (f5) {$\grho,\re$};
      \vertex[above right = of v2] (f6) {$\onu,\g$};
      \vertex [left = 0.5em of v1] {$G^{\g\re}_{\g\re}$};
      \vertex [right = 0.5em of v2] {$G^{\re\g}_{\g\re}$};
      \diagram* {
        (f1) --[fermion, color = lightblue] (v1)
        --[fermion, color = lightblue] (f3),
        (f2) --[fermion, color = orange] (v1)
        --[charged scalar,
        edge label = {$\mathcolor{black}{n}
          \mathcolor{orange}{\nu}\mathcolor{black}{,\re}$},
        color = orange] (v2),
        (f4) --[fermion, color = green] (v2)
        --[fermion, color = green] (f5),
        (v2) --[fermion, color = orange] (f6), };
    \end{feynman}
  \end{tikzpicture}
  \begin{array}{ll}
    &= \P_0 \p{\f12 K_n G^{\re\g}_{\g\re}
      \c_{\onu,\g}^\dag \c_{\grho,\re}^\dag
      \c_{\grho,\g} \c_{n\onu,\re}}
      \f1{E_n}
      \p{\f12 K_n G^{\g\re}_{\g\re}
      \c_{\bmu,\g}^\dag \c_{n\onu,\re}^\dag
      \c_{\onu,\re} \c_{\bmu,\g}} \P_0 \\[1em]
    &= \f14 \f{K_n^2}{E_n}~ G^{\g\re}_{\g\re} G^{\re\g}_{\g\re}~
      \P_0 \c_{\bmu,\g}^\dag \c_{\onu,\g}^\dag \c_{\grho,\re}^\dag
      \c_{\grho,\g} \c_{\onu,\re} \c_{\bmu,\g} \P_0
  \end{array}
  \)
  \caption{\footnotesize An example second-order diagram and the
    corresponding three-body interaction term in $H_{\t{int}}^{(2)}$,
    with $n>0$ and $\c_{\mu s}\equiv\c_{0,\mu s}$.  Diagrams are read
    from left to right to construct a sequence of operators from right
    to left.  Solid (dashed) lines represent field operators acting on
    the lowest (arbitrary) motional states.  For the sake of
    presentation, this diagram has colors associated with nuclear spin
    and orbital states, an arrow on each line to emphasize that they
    are directed left-to-right, and an explicit coupling constant
    written next to each vertex; we will generally not include these
    features, as they are not necessary to uniquely identify the term
    represented by a diagram.  We will also drop explicit appearances
    of the ground-state projector $\P_0$ in our expressions, with the
    understanding that the low-energy effective theory implicitly
    addresses only non-interacting motional ground states.}
  \label{fig:diagram}
\end{figure}

The form of the bare two-body interaction Hamiltonian $H_{\t{int}}$ in
\eqref{eq:H_int} motivates a diagrammatic representation of terms in
the effective Hamiltonian $H_{\t{eff}}$ in \eqref{eq:H_int_eff},
similarly to the diagrams used to represent elements of the scattering
matrix in standard quantum field theory.  Effective $M$-body
interaction terms at order $p$ in $H_{\t{int}}$ can be represented by
directed graphs containing $p$ vertices with degree greater than one,
which we call {\it internal vervices}.  Each internal vertex and its
associated edges correspond respectively to a coupling constant and
the associated field operators in $H_{\t{int}}$.  An example 2-vertex
diagram representing an effective 3-body interaction term is provided
in figure \ref{fig:diagram}.  All diagrams are read from left to right
to construct a sequence of operators from right to left; the internal
vertices of a diagram are thus strictly ordered, with the $n$-th
internal vertex from the left corresponding to the $n$-th interaction
Hamiltonian $H_{\t{int}}$ from the right in \eqref{eq:H_int_1_2} or
\eqref{eq:H_int_3}.  Solid (dashed) lines represent field operators
acting on the lowest (arbitrary) motional states.  Spatial overlap
factors at each vertex are determined by the motional states of the
edges which connect to (i.e.~field operators associated with) that
vertex.  While it is possible to construct explicit rules for
determining the energetic suppression factors (i.e.~from $\I$) of the
term represented by a diagram, these factors are most easily
determined by examination of the effective Hamiltonians in
\eqref{eq:H_int_1_2} and \eqref{eq:H_int_3}.

The diagram in figure \ref{fig:diagram} explicitly labels all edges
with indices of the corresponding field operators, but in general we
may suppress these indices, in which case the diagram includes a sum
over the suppressed indices.  These sums are performed over all
allowed values of the suppressed indices, with the restriction that
virtual states (i.e.~vertical slices of the diagram between internal
vertices) represented with dashed lines must have at least one
motional excitation.  While we include factors of $1/2$ from
$H_{\t{int}}$ as expressed in \eqref{eq:H_int} in the definition of a
diagram, in all but the two-body case these factors of $1/2$ will be
cancelled out by corresponding symmetry factors, i.e.~the appearance
of duplicate diagrams which are equal up to a relabeling of indices
(see Appendix \ref{sec:diagrams}).  The explicit signs and factor of
$1/2$ which appear in the effective Hamiltonians in
\eqref{eq:H_int_1_2} and \eqref{eq:H_int_3} are not included in the
definition of a diagram, and must be kept track of manually.

\subsection{Effective two-body interactions and renormalization}
\label{sec:two_body}

The effective two-body Hamiltonian in \eqref{eq:H_int_eff} has
contributions at all orders in the coupling constants, and can be
expanded in the form
\begin{align}
  H_2 = \shrink{
    \begin{tikzpicture}
      \begin{feynman}
        \vertex (v);
        \vertex[above left = of v] (f1);
        \vertex[below left = of v] (f2);
        \vertex[above right = of v] (f3);
        \vertex[below right = of v] (f4);
        \diagram* {
          (f1) -- (v),
          (f2) -- (v),
          (v) -- (f3),
          (v) -- (f4) };
      \end{feynman}
    \end{tikzpicture}}
  - \shrink{
    \begin{tikzpicture}
      \begin{feynman}
        \vertex (v1);
        \vertex[above left = of v1] (f1);
        \vertex[below left = of v1] (f2);
        \vertex[right = of v1] (v2);
        \vertex[above right = of v2] (f3);
        \vertex[below right = of v2] (f4);
        \diagram* {
          (f1) -- (v1),
          (f2) -- (v1),
          (v2) -- (f3),
          (v2) -- (f4),
          (v1) --[scalar, half left] (v2),
          (v1) --[scalar, half right] (v2) };
      \end{feynman}
    \end{tikzpicture}}
  + \shrink{
    \begin{tikzpicture}
      \begin{feynman}
        \vertex (v1);
        \vertex[above left = of v1] (f1);
        \vertex[below left = of v1] (f2);
        \vertex[right = of v1] (v2);
        \vertex[right = of v2] (v3);
        \vertex[above right = of v3] (f3);
        \vertex[below right = of v3] (f4);
        \diagram* {
          (f1) -- (v1),
          (f2) -- (v1),
          (v3) -- (f3),
          (v3) -- (f4),
          (v1)
          --[scalar, half left] (v2)
          --[scalar, half left] (v3),
          (v1)
          --[scalar, half right] (v2)
          --[scalar, half right] (v3) };
      \end{feynman}
    \end{tikzpicture}}
  + \cdots \equiv \shrink{
    \begin{tikzpicture}
      \begin{feynman}
        \vertex[blob] (v) {};
        \vertex[above left = of v] (f1);
        \vertex[below left = of v] (f2);
        \vertex[above right = of v] (f3);
        \vertex[below right = of v] (f4);
        \diagram* {
          (f1) -- (v),
          (f2) -- (v),
          (v) -- (f3),
          (v) -- (f4) };
      \end{feynman}
    \end{tikzpicture}},
  \label{eq:H_2_expansion}
\end{align}
where the blob on the right schematically represents the net effective
two-body interaction.  On physical grounds, the net two-body
interaction must clearly be finite, but individual sums over excited
states in the loop diagrams of \eqref{eq:H_2_expansion} may generally
diverge \cite{johnson2012effective}.  These divergences ultimately
appear due to our use of effective field theory to describe
inter-atomic interactions in \eqref{eq:H_int_start},
\eqref{eq:H_int_fields}, and \eqref{eq:H_int}, rather than a detailed
microscopic description of two-atom scattering.  Divergences of this
sort are a generic feature of field theories, and can be dealt with
using standard techniques such as renormalization.  We therefore
renormalize our coupling constants by introducing counter-terms
$\widetilde G_X$ into the interaction Hamiltonian.

The introduction of counter-terms is merely a formal decomposition of
the ``bare'' coupling constants $G_X^{\t{bare}}$ that are used in
\eqref{eq:H_2_expansion} as $G_X^{\t{bare}} = G_X + \widetilde G_X$.
In performing such a decomposition, we are free to choose the values
of $G_X$, which in turn fixes the values of
$\widetilde G_X \equiv G_X^{\t{bare}} - G_X$.  For convenience, we can
choose the values of $G_X$ to be those of the net effective coupling
constants on the right-hand side of \eqref{eq:H_2_expansion}.
Representing the new coupling constants $G_X$ by regular vertices and
the counter-terms $\widetilde G_X$ by a crossed dot (i.e.~$\otimes$),
this choice leads to the {\it renormalization condition}
\begin{align}
  \substack{
    \shrink{
      \begin{tikzpicture}
        \begin{feynman}
          \vertex (v);
          \vertex[above left = of v] (f1);
          \vertex[below left = of v] (f2);
          \vertex[above right = of v] (f3);
          \vertex[below right = of v] (f4);
          \diagram* {
            (f1) -- (v),
            (f2) -- (v),
            (v) -- (f3),
            (v) -- (f4) };
        \end{feynman}
      \end{tikzpicture}}
    \\ \O(G)}
  + \substack{
    \shrink{
      \begin{tikzpicture}
        \begin{feynman}
          \vertex[crossed dot] (v) {};
          \vertex[above left = of v] (f1);
          \vertex[below left = of v] (f2);
          \vertex[above right = of v] (f3);
          \vertex[below right = of v] (f4);
          \diagram* {
            (f1) -- (v),
            (f2) -- (v),
            (v) -- (f3),
            (v) -- (f4) };
        \end{feynman}
      \end{tikzpicture}}
    \\ \O(\widetilde G)}
  - \substack{
    \shrink{
      \begin{tikzpicture}
        \begin{feynman}
          \vertex (v1);
          \vertex[above left = of v1] (f1);
          \vertex[below left = of v1] (f2);
          \vertex[right = of v1] (v2);
          \vertex[above right = of v2] (f3);
          \vertex[below right = of v2] (f4);
          \diagram* {
            (f1) -- (v1),
            (f2) -- (v1),
            (v2) -- (f3),
            (v2) -- (f4),
            (v1) --[scalar, half left] (v2),
            (v1) --[scalar, half right] (v2) };
        \end{feynman}
      \end{tikzpicture}}
    \\ \O(G^2)}
  + \cdots = \substack{
    \shrink{
      \begin{tikzpicture}
        \begin{feynman}
          \vertex (v);
          \vertex[above left = of v] (f1);
          \vertex[below left = of v] (f2);
          \vertex[above right = of v] (f3);
          \vertex[below right = of v] (f4);
          \diagram* {
            (f1) -- (v),
            (f2) -- (v),
            (v) -- (f3),
            (v) -- (f4) };
        \end{feynman}
      \end{tikzpicture}}
    \\ \O(G)}.
  \label{eq:counter_term}
\end{align}
This renormalization condition has the benefit of allowing us to
express effective two-body interactions simply in terms of net
effective two-body coupling constants, rather than in terms of long
sums at all order of the bare coupling constants.  By construction,
the counter-terms we have introduced exactly cancel all terms beyond
leading order in \eqref{eq:H_2_expansion}, which implies that the
effective two-body interaction Hamiltonian is simply
\begin{align}
  H_2 = \shrink{
    \begin{tikzpicture}
      \begin{feynman}
        \vertex (v);
        \vertex[above left = of v] (f1);
        \vertex[below left = of v] (f2);
        \vertex[above right = of v] (f3);
        \vertex[below right = of v] (f4);
        \diagram* {
          (f1) -- (v),
          (f2) -- (v),
          (v) -- (f3),
          (v) -- (f4), };
      \end{feynman}
    \end{tikzpicture}}
  = \f12 \alpha_2^{(1)} \sum_{\abs{\set{\mu,\nu}}=2} G^{qr}_{st}
  \c_{\mu s}^\dag \c_{\nu t}^\dag \c_{\nu r} \c_{\mu q},
  \label{eq:H_2}
\end{align}
where for consistency with existing literature
\cite{johnson2012effective} we define $\alpha_2^{(1)} \equiv K$ as the
overlap integral between two atoms occupying non-interacting motional
ground states.

Before moving on to consider effective three-body interactions, there
are a few comments we must make concerning renormalization and the
result in \eqref{eq:H_2}.  First, the effective two-body interaction
$H_2$ in \eqref{eq:H_2} takes the same form as the bare two-body
interaction $H_{\t{int}}$ in \eqref{eq:H_int}, but without excited
motional states, and with renormalized coupling constants.  Our choice
of renormalization scheme, while convenient for the analytical
development of a low-energy effective theory, no longer allows us to
use the coupling constants $G_X$ as defined by the free-space
scattering lengths $a_X$ in \eqref{eq:couplings} to compute
interaction energies.  The renormalization condition in
\eqref{eq:counter_term} explicitly fixes $G_X$ to the net effective
coupling constants in any given setting.  Instead of using free-space
coupling constants to compute interaction energies in a lattice, we
must therefore first compute the effective coupling constants
$G_X^{\t{lattice}}\p{\U}$, which now depend on the lattice depth $\U$,
and in turn use these effective coupling constants to compute
interaction energies.  We discuss the calculation of effective
coupling constants in Appendix \ref{sec:renormalization}.

Second, the renormalization condition in \eqref{eq:counter_term}
implies that the counter terms $\widetilde G_X$ are second order in
the coupling constants $G_X$, i.e.~$\widetilde G_X\sim G_X^2$.
Although the effective Hamiltonian expansions in \eqref{eq:H_int_1_2}
and \eqref{eq:H_int_3} are organized in powers of the interaction
Hamiltonian $H_{\t{int}}$, the couplings $G_X$ are the ``small''
parameters in which we can formally organize a perturbation theory;
more specifically, the formally small quantities organizing our
perturbation theory are two-body ground-state interaction energies
(proportional to the couplings $G_X$) divided by the spectral gap of
the single-atom Hamiltonian $H_0$ (see Appendix
\ref{sec:pert_params}).  If $M$ atoms can only couple through terms
represented by a $p$-vertex diagrams for $p\ge p_M^{\t{min}}$, then
the leading order contribution to $M$-body interactions is order
$p_M^{\t{min}}$ in the couplings $G_X$.  If the same
$p_M^{\t{min}}$-vertex diagrams involve any counter-terms, however,
then these diagrams are at least order $p_M^{\t{min}}+1$ in the
couplings $G_X$.  Counter-terms therefore only appear at
next-to-leading order (NLO) in the calculation of effective $M$-body
interactions.

Finally, our result in \eqref{eq:H_2} neglects the effect of
momentum-dependent two-body scattering.  When the effective range of
interactions is comparable to the scattering lengths $a_X$, as is the
case for ultracold ${}^{87}$Sr, these momentum-dependent effects are
third order in the coupling constants $G_X$.  Just as the $\O\p{G^2}$
counter-terms do not affect $M$-body interactions until
next-to-leading order (NLO), the $\O\p{G^3}$ momentum-dependent terms
do not come into play until next-next-leading order (NNLO).  Given
that we develop our low-energy effective theory through third order in
the coupling constants, these interactions will not appear in any of
our three- and four-body calculations, but they do need to be
considered in the calculation of pair-wise interaction energies.  The
primary interest of our work, however, concerns effective $M$-body
interactions for $M\ge3$; we therefore defer the calculation of
momentum-dependent two-body interactions to Appendix
\ref{sec:momentum_dependence}.

\subsection{Effective three-body interactions at second order}

Our theory of effective multi-body interactions assumes no
non-universal contribution to the three-body interaction energy, which
is to say that we assume the absence of real (as opposed to
effective), bare three-body interactions.  Consequently, three-body
interactions do not appear until second order in the coupling
constants of the effective theory, in the expansion of
$H_{\t{int}}^{(2)}$ in \eqref{eq:H_int_1_2}.  The virtual state of
three-body terms in $H_{\t{int}}^{(2)}$ cannot have two atoms in
excited motional states, as otherwise the second application of
$H_{\t{int}}$ in $H_{\t{int}}^{(2)}$ would have to address both of
theses atoms to bring them back down to the ground state, resulting in
a two-body process as in the second diagram of
\eqref{eq:H_2_expansion}.  All second-order three-body terms must
therefore have only one excited atom in the virtual state, and take
the form
\begin{align}
  \begin{tikzpicture}
    \begin{feynman}
      \vertex (v1);
      \vertex[above left = of v1] (f1) {$\mu r$};
      \vertex[below left = of v1] (f2) {$\nu s$};
      \vertex[right = of v1] (vm);
      \vertex[right = of vm] (v2);
      \vertex[above = of vm] (f3) {$\mu r'$};
      \vertex[below = of vm] (f4) {$\rho t$};
      \vertex[below right = of v2] (f5) {$\rho t'$};
      \vertex[above right = of v2] (f6) {$\nu s''$};
      \diagram* {
        (f1) -- (v1) -- (f3),
        (f2) -- (v1) --  [scalar, edge label=$n\nu s'$] (v2),
        (f4) -- (v2) -- (f5),
        (v2) -- (f6), };
    \end{feynman}
  \end{tikzpicture}
  \propto K_n^2 G^{rs}_{r's'} G^{s't}_{s''t'}
  \c_{\mu r'}^\dag \c_{\nu s''}^\dag \c_{\rho t'}^\dag
  \c_{\rho t} \c_{\nu s} \c_{\mu r}
  \label{eq:H_3_2_diagram}.
\end{align}
Unlike for the two-body diagram in \eqref{eq:H_2}, the explicit
factors of $1/2$ which appear in the bare two-body Hamiltonian
$H_{\t{int}}$ in \eqref{eq:H_int} are now cancelled out by symmetry
factors which account for duplicate diagrams; this cancellation will
generally occur for all connected $M$-body diagrams with $M>2$ (see
Appendix \ref{sec:diagrams}).  The net effective three-body
interaction Hamiltonian at second order is then given by the sum over
all diagrams of the form in \eqref{eq:H_3_2_diagram}, i.e.
\begin{align}
  H_3^{(2)} = - \shrink{
    \begin{tikzpicture}
      \begin{feynman}
        \vertex (v1);
        \vertex[above left = of v1] (f1);
        \vertex[below left = of v1] (f2);
        \vertex[right = 4em of v1] (v2);
        \vertex[above right = of v1] (f3);
        \vertex[below left = of v2] (f4);
        \vertex[below right = of v2] (f5);
        \vertex[above right = of v2] (f6);
        \diagram* {
          (f1) -- (v1) -- (f3),
          (f2) -- (v1) --[scalar] (v2),
          (f4) -- (v2) -- (f5),
          (v2) -- (f6), };
      \end{feynman}
    \end{tikzpicture}}
  = -\alpha_3^{(2)} \sum_{\abs{\set{\mu,\nu,\rho}}=3}
  G^{rs}_{r's'} G^{s't}_{s''t'}
  \c_{\mu r'}^\dag \c_{\nu s''}^\dag \c_{\rho t'}^\dag
  \c_{\rho t} \c_{\nu s} \c_{\mu r},
  \label{eq:H_3_2}
\end{align}
where $\alpha_3^{(2)} \equiv \sum_{n>0} K_n^2/E_n$, and the preceding
minus sign is as prescribed by $H_{\t{int}}^{(2)}$ in
\eqref{eq:H_int_1_2}.

\subsection{Effective three-body interactions at third order}

The third-order effective interaction Hamiltonian $H_{\t{int}}^{(3)}$
in \eqref{eq:H_int_3} contains both three- and four-body terms.  To
compactly enumerate and evaluate all three-body diagrams at third
order, we introduce an expanded coupling symbol
\begin{align}
  G^{\mu q;\nu r}_{\rho s;\sigma t} \equiv \left\{
    \begin{array}{ll}
      G^{qr}_{st} & ~ \p{\mu,\nu} = \p{\rho,\sigma}
      \\
      - G^{qr}_{ts} & ~ \p{\mu,\nu} = \p{\sigma,\rho}
      \\
      0 & ~ \t{otherwise}
    \end{array}\right.
  \label{eq:G_expanded}
\end{align}
for more general
$\p{\mu,q}+\p{\nu,r}\leftrightarrow\p{\rho,s}+\p{\sigma,t}$ coupling
induced by terms proportional to
$\c_{\rho s}^\dag \c_{\sigma t}^\dag \c_{\nu r} \c_{\mu q}$.  The
minus sign in \eqref{eq:G_expanded} accounts for fermionic statistics:
if $\p{\mu,\nu}=\p{\sigma,\rho}$, then we are considering a term of
the form
\begin{align}
  G^{\mu q;\nu r}_{\nu s;\mu t}
  \c_{\nu s}^\dag \c_{\mu t}^\dag \c_{\nu r} \c_{\mu q}
  = -G^{qr}_{ts} \c_{\nu s}^\dag \c_{\mu t}^\dag \c_{\nu r} \c_{\mu q}
  = G^{qr}_{ts} \c_{\mu t}^\dag \c_{\nu s}^\dag \c_{\nu r} \c_{\mu q}.
\end{align}
At the cost of introducing an additional sum over new nuclear spin
indices, the expanded coupling symbol allows us to collect together
diagrams which have the same graph topology, but represent different
matrix elements of the effective Hamiltonian due to the exchange of
nuclear spins at a vertex.  The third order three-body diagrams in
$H_{\t{int}}^{(3)}$ are then
\begin{align}
  \begin{tikzpicture}
    \begin{feynman}
      \vertex (v1);
      \vertex[below right = of v1] (v2);
      \vertex[above right = of v2] (v3);
      \vertex[above left = of v1] (f1) {$\mu r$};
      \vertex[left = of v1] (f2) {$\nu s$};
      \vertex[below left = of v2] (f3) {$\rho t$};
      \vertex[above right = of v3] (f4) {$\mu r''$};
      \vertex[right = of v3] (f5) {$\nu's'''$};
      \vertex[below right = of v2] (f6) {$\rho't'$};
      \diagram* {
        (f1) -- (v1) --[scalar, edge label=$\ell\mu r'$] (v3) -- (f4),
        (f2) -- (v1)
        --[scalar, edge label'=$m\nu s'$] (v2)
        --[scalar, edge label'=$n\nu's''$] (v3)
        -- (f5),
        (f3) -- (v2) -- (f6), };
    \end{feynman}
  \end{tikzpicture}
  \propto K_{\ell m} K^m_n K_{\ell n}
  G^{rs}_{r's'} G^{\nu s';\rho t}_{\nu's'';\rho't'} G^{r's''}_{r''s'''}
  \c_{\mu r''}^\dag \c_{\nu's'''}^\dag \c_{\rho't'}^\dag
  \c_{\rho t} \c_{\nu s} \c_{\mu r},
  \label{eq:H_3_3_S}
\end{align}
\begin{align}
  \begin{tikzpicture}
    \begin{feynman}
      \vertex (v1);
      \vertex[right = of v1] (v2);
      \vertex[below right = of v2] (v3);
      \vertex[above left = of v1] (f1) {$\mu r$};
      \vertex[below left = of v1] (f2) {$\nu s$};
      \vertex[below left = of v3] (f3) {$\rho t$};
      \vertex[above right = of v2] (f4) {$\mu r''$};
      \vertex[above right = of v3] (f5) {$\nu s'''$};
      \vertex[below right = of v3] (f6) {$\rho t'$};
      \diagram* {
        (f1) -- (v1)
        --[half left, scalar, edge label=$\ell\mu r'$] (v2)
        -- (f4),
        (f2) -- (v1)
        --[half right, scalar, edge label'=$m\nu s'$] (v2)
        --[scalar, edge label=$n\nu s''$] (v3)
        -- (f5),
        (f3) -- (v3) -- (f6), };
    \end{feynman}
  \end{tikzpicture}
  \propto K_{\ell m} K^{\ell m}_n K_n
  G^{rs}_{r's'} G^{r's'}_{r''s''} G^{s''t}_{s'''t'}
  \c_{\mu r''}^\dag \c_{\nu s'''}^\dag \c_{\rho t'}^\dag
  \c_{\rho t} \c_{\nu s} \c_{\mu r},
  \label{eq:H_3_3_O}
\end{align}
and the mirror image of \eqref{eq:H_3_3_O}.  As prescribed by
$H_{\t{int}}^{(3)}$ in \eqref{eq:H_int_3}, these diagrams have an
associated minus sign if they contain only one excited virtual state,
and a factor of $1/2$ if they contain a virtual ground state.
Remembering that counter-terms are $\O\p{G^2}$, there are additionally
two third-order three-body diagrams in $H_{\t{int}}^{(2)}$, namely
\begin{align}
  \begin{tikzpicture}
    \begin{feynman}
      \vertex[crossed dot] (v1) {};
      \vertex[above left = 4em of v1] (f1) {$\mu r$};
      \vertex[below left = 4em of v1] (f2) {$\nu s$};
      \vertex[right = 4em of v1] (v2);
      \vertex[above right = 4em of v1] (f3) {$\mu r'$};
      \vertex[below left = of v2] (f4) {$\rho t$};
      \vertex[below right = of v2] (f5) {$\rho t'$};
      \vertex[above right = of v2] (f6) {$\nu s''$};
      \diagram* {
        (f1) -- (v1) -- (f3),
        (f2) -- (v1)
        --[scalar, edge label=$n\nu s'$] (v2),
        (f4) -- (v2) -- (f5),
        (v2) -- (f6), };
    \end{feynman}
  \end{tikzpicture}
  \propto K_n^2 \widetilde G^{rs}_{r's'} G^{s't}_{s''t'}
  \c_{\mu r'}^\dag \c_{\nu s''}^\dag \c_{\rho t'}^\dag
  \c_{\rho t} \c_{\nu s} \c_{\mu r}
  \label{eq:H_3_3_counter_term}
\end{align}
and its mirror image, where $\widetilde G^{qr}_{st}$ is equal to the
counter-term associated with $G^{qr}_{st}$.

The net contribution to the third-order three-body interaction
Hamiltonian from three-particle-loop diagrams of the form in
\eqref{eq:H_3_3_S} is
\begin{align}
  \shrink{
    \begin{tikzpicture}
      \begin{feynman}
        \vertex (v1);
        \vertex[below right = of v1] (v2);
        \vertex[above right = of v2] (v3);
        \vertex[above left = of v1] (f1);
        \vertex[left = of v1] (f2);
        \vertex[below left = of v2] (f3);
        \vertex[above right = of v3] (f4);
        \vertex[right = of v3] (f5);
        \vertex[below right = of v2] (f6);
        \diagram* {
          (f1) -- (v1) --[scalar] (v3) -- (f4),
          (f2) -- (v1) --[scalar] (v2) --[scalar] (v3) -- (f5),
          (f3) -- (v2) -- (f6), };
      \end{feynman}
    \end{tikzpicture}}
  - \f12 \shrink{
    \begin{tikzpicture}
      \begin{feynman}
        \vertex (v1);
        \vertex[below right = of v1] (v2);
        \vertex[above right = of v2] (v3);
        \vertex[above left = of v1] (f1);
        \vertex[left = of v1] (f2);
        \vertex[below left = of v2] (f3);
        \vertex[above right = of v3] (f4);
        \vertex[right = of v3] (f5);
        \vertex[below right = of v2] (f6);
        \diagram* {
          (f1) -- (v1) -- (v3) -- (f4),
          (f2) -- (v1) --[scalar] (v2) -- (v3) -- (f5),
          (f3) -- (v2) -- (f6), };
      \end{feynman}
    \end{tikzpicture}}
  - \f12 \shrink{
    \begin{tikzpicture}
      \begin{feynman}
        \vertex (v1);
        \vertex[below right = of v1] (v2);
        \vertex[above right = of v2] (v3);
        \vertex[above left = of v1] (f1);
        \vertex[left = of v1] (f2);
        \vertex[below left = of v2] (f3);
        \vertex[above right = of v3] (f4);
        \vertex[right = of v3] (f5);
        \vertex[below right = of v2] (f6);
        \diagram* {
          (f1) -- (v1) -- (v3) -- (f4),
          (f2) -- (v1) -- (v2) --[scalar] (v3) -- (f5),
          (f3) -- (v2) -- (f6), };
      \end{feynman}
    \end{tikzpicture}}
  = \p{\alpha_{3,1}^{(3)} - \alpha_5^{(3)}} \H_{3,1}^{(3)},
\end{align}
where
\begin{align}
  \alpha_{3,1}^{(3)} \equiv \sum_{\substack{\ell+m>0\\\ell+n>0}}
  \f{K_{\ell m} K^m_n K_{\ell n}}{E_{\ell m} E_{\ell n}},
  &&
  \alpha_5^{(3)}
  \equiv  K \sum_{n>0} \f{K_n^2}{E_n^2},
  \label{eq:alph_3_3_1}
\end{align}
and
\begin{align}
  \H_{3,1}^{(3)} \equiv \sum_{\abs{\set{\mu,\nu,\rho}}=3}
  G^{rs}_{r's'} G^{\nu s'\rho t}_{\nu's''\rho't'} G^{r's''}_{r''s'''}
  \c_{\mu r''}^\dag \c_{\nu's'''}^\dag \c_{\rho't'}^\dag
  \c_{\rho t} \c_{\nu s} \c_{\mu r}.
\end{align}
Even though this contribution comes from loop diagrams, the factors
$\alpha_{3,1}^{(3)}$ and $\alpha_5^{(3)}$ in \eqref{eq:alph_3_3_1} are
finite.  At large motional state indices $n$, atoms become free
particles for which $n$ essentially indexes discrete momentum states.
These atoms thus have an energy which asymptotically scales as
$E_n\sim n^2\equiv n_\x^2+n_\y^2+n_\z^2$.  Furthermore, the
oscillatory behavior of atomic wavefunctions with increasing motional
state indices $\ell,m$ implies that the overlap integral $K_{\ell m}$
becomes sharply peaked at $\ell\approx m$ as $\ell$ and $m$ get large.
The asymptotic behavior of $\alpha_{3,1}^{(3)}$ at large $\ell,m,n$ is
therefore
\begin{align}
  \alpha_{3,1}^{(3)}
  \sim \int \f{\d^3\ell~\d^3m~\d^3n}{\p{\ell^2+m^2}\p{\ell^2+n^2}}
  ~\delta\p{\ell-m}\delta\p{\ell-n}
  \sim \int \f{\d^3\ell}{\ell^4}
  \sim \int_{\ell_{\t{min}}}^\infty \f{\d\ell}{\ell^2}
  \sim \f1{\ell_{\t{min}}},
  \label{eq:alph_3_3_1_integral}
\end{align}
where in the last integral we changed to spherical coordinates, and
$\ell_{\t{min}}^2$ is the minimum value of $\ell^2$ for which
\begin{enumerate*}
\item the energy $E_\ell\sim\ell^2$, and
\item the integral expression in \eqref{eq:alph_3_3_1_integral} is a
  good approximation to the corresponding sum in
  \eqref{eq:alph_3_3_1}.
\end{enumerate*}
Note that the introduction of $\ell_{\t{min}}$ amounts to neglecting a
finite number of terms in the sum over $\ell,m,n$ in
\eqref{eq:alph_3_3_1}, whose contribution to the value of
$\alpha_{3,1}^{(3)}$ is finite.  Convergence of $\alpha_5^{(3)}$ is
similarly guaranteed by the fact that the overlap integral $K_n$ does
not asymptotically grow with increasing $n$, such that
$\alpha_5^{(3)}$ asymptotically behaves as
\begin{align}
  \alpha_5^{(3)} \sim \int \f{\d^3 n}{n^4}
  \sim \int_{n_{\t{min}}}^\infty \f{\d n}{n^2}
  \sim \f1{n_{\t{min}}},
  \label{eq:alph_5_3_integral}
\end{align}
where again $n_{\t{min}}$ is defined similarly to $\ell_{\t{min}}$.

The sum over loop diagrams in \eqref{eq:H_3_3_O}, meanwhile, contains
a divergence that must be cancelled out by the counter-terms in
\eqref{eq:H_3_3_counter_term}.  To leading order in the coupling
constants, the renormalization condition in \eqref{eq:counter_term}
implies that
\begin{align}
  \shrink{
    \begin{tikzpicture}
      \begin{feynman}
        \vertex[crossed dot] (v) {};
        \vertex[above left = of v] (f1);
        \vertex[below left = of v] (f2);
        \vertex[above right = of v] (f3);
        \vertex[below right = of v] (f4);
        \diagram* {
          (f1) -- (v),
          (f2) -- (v),
          (v) -- (f3),
          (v) -- (f4), };
      \end{feynman}
    \end{tikzpicture}}
  = \shrink{
    \begin{tikzpicture}
      \begin{feynman}
        \vertex (v1);
        \vertex[right = of v1] (v2);
        \vertex[above left = of v1] (f1);
        \vertex[below left = of v1] (f2);
        \vertex[above right = of v2] (f3);
        \vertex[below right = of v2] (f4);
        \diagram* {
          (f1) -- (v1) --[half left, scalar] (v2) -- (f3),
          (f2) -- (v1) --[half right, scalar] (v2) -- (f4), };
      \end{feynman}
    \end{tikzpicture}},
\end{align}
which can be expanded to find
\begin{align}
  K \widetilde G^{rs}_{r''s''}
  = \sum_{\ell,m,r',s'} \f{K_{\ell m}^2}{E_{\ell m}}
  G^{rs}_{r's'} G^{r's'}_{r''s''}.
\end{align}
In terms of ordinary coupling constants, the counter-term diagram in
\eqref{eq:H_3_3_counter_term} is therefore
\begin{align}
  \begin{tikzpicture}
    \begin{feynman}
      \vertex[crossed dot] (v1) {};
      \vertex[above left = 4em of v1] (f1) {$\mu r$};
      \vertex[below left = 4em of v1] (f2) {$\nu s$};
      \vertex[right = 4em of v1] (v2);
      \vertex[above right = 4em of v1] (f3) {$\mu r''$};
      \vertex[below left = of v2] (f4) {$\rho t$};
      \vertex[below right = of v2] (f5) {$\rho t'$};
      \vertex[above right = of v2] (f6) {$\nu s'''$};
      \diagram* {
        (f1) -- (v1) -- (f3),
        (f2) -- (v1)
        --[scalar, edge label=$n\nu s''$] (v2),
        (f4) -- (v2) -- (f5),
        (v2) -- (f6), };
    \end{feynman}
  \end{tikzpicture}
  = \sum_{\ell,m,r',s'}
  \f{K_{\ell m}^2 K_n^2}{K E_{\ell m} E_n}
  G^{rs}_{r's'} G^{r's'}_{r''s''} G^{s''t}_{s'''t'}
  \c_{\mu r''}^\dag \c_{\nu s'''}^\dag \c_{\rho t'}^\dag
  \c_{\rho t} \c_{\nu s} \c_{\mu r}.
\end{align}
Altogether, the contribution to the third-order three-body interaction
Hamiltonian from loop diagrams of the form in \eqref{eq:H_3_3_O} and
counter-term diagrams of the form in \eqref{eq:H_3_3_counter_term} is
\begin{multline}
  \shrink{
    \begin{tikzpicture}
      \begin{feynman}
        \vertex (v1);
        \vertex[right = of v1] (v2);
        \vertex[below right = of v2] (v3);
        \vertex[above left = of v1] (f1);
        \vertex[below left = of v1] (f2);
        \vertex[below left = of v3] (f3);
        \vertex[above right = of v2] (f4);
        \vertex[above right = of v3] (f5);
        \vertex[below right = of v3] (f6);
        \diagram* {
          (f1) -- (v1) --[half left, scalar] (v2) -- (f4),
          (f2) -- (v1) --[half right, scalar] (v2)
          --[scalar] (v3) -- (f5),
          (f3) -- (v3) -- (f6), };
      \end{feynman}
    \end{tikzpicture}}
    - \shrink{
    \begin{tikzpicture}
      \begin{feynman}
        \vertex[crossed dot] (v1) {};
        \vertex[right = 4em of v1] (v2);
        \vertex[above left = of v1] (f1);
        \vertex[below left = of v1] (f2);
        \vertex[above right = of v1] (f3);
        \vertex[below left = of v2] (f4);
        \vertex[below right = of v2] (f5);
        \vertex[above right = of v2] (f6);
        \diagram* {
          (f1) -- (v1) -- (f3),
          (f2) -- (v1) --[scalar] (v2),
          (f4) -- (v2) -- (f5),
          (v2) -- (f6), };
      \end{feynman}
    \end{tikzpicture}}
  - \f12 \shrink{
    \begin{tikzpicture}
      \begin{feynman}
        \vertex (v1);
        \vertex[right = of v1] (v2);
        \vertex[below right = of v2] (v3);
        \vertex[above left = of v1] (f1);
        \vertex[below left = of v1] (f2);
        \vertex[below left = of v3] (f3);
        \vertex[above right = of v2] (f4);
        \vertex[above right = of v3] (f5);
        \vertex[below right = of v3] (f6);
        \diagram* {
          (f1) -- (v1) --[half left, scalar] (v2) -- (f4),
          (f2) -- (v1) --[half right, scalar] (v2)
          -- (v3) -- (f5),
          (f3) -- (v3) -- (f6), };
      \end{feynman}
    \end{tikzpicture}}
  - \f12 \shrink{
    \begin{tikzpicture}
      \begin{feynman}
        \vertex (v1);
        \vertex[right = of v1] (v2);
        \vertex[below right = of v2] (v3);
        \vertex[above left = of v1] (f1);
        \vertex[below left = of v1] (f2);
        \vertex[below left = of v3] (f3);
        \vertex[above right = of v2] (f4);
        \vertex[above right = of v3] (f5);
        \vertex[below right = of v3] (f6);
        \diagram* {
          (f1) -- (v1) --[half left] (v2) -- (f4),
          (f2) -- (v1) --[half right] (v2)
          --[scalar] (v3) -- (f5),
          (f3) -- (v3) -- (f6), };
      \end{feynman}
    \end{tikzpicture}} \\
  = \p{\alpha_{3,2}^{(3)} - \f12\alpha_{4,3}^{(3)} - \f12\alpha_5^{(3)}}
  \H_{3,2}^{(3)},
\end{multline}
where
\begin{align}
  \alpha_{3,2}^{(3)}
  \equiv \sum_{\substack{\ell+m>0\\n>0}}
  \f{K_{\ell m} K_n}{E_{\ell m} E_n}
  \p{K^{\ell m}_n - \f{K_{\ell m} K_n}{K}},
  &&
  \alpha_{4,3}^{(3)}
  \equiv K \sum_{m+n>0} \f{K_{mn}^2}{E_{mn}^2},
  \label{eq:alph_3_3_2}
\end{align}
and
\begin{align}
  \H_{3,2}^{(3)} \equiv \sum_{\abs{\set{\mu,\nu,\rho}}=3}
  G^{rs}_{r's'} G^{r's'}_{r''s''} G^{s''t}_{s'''t'}
  \c_{\mu r''}^\dag \c_{\nu s'''}^\dag \c_{\rho t'}^\dag
  \c_{\rho t} \c_{\nu s} \c_{\mu r}.
\end{align}
An equal contribution comes from the mirror images of these diagrams,
such that the net third-order three-body interaction Hamiltonian is
\begin{align}
  H_3^{(3)} = \p{\alpha_{3,1}^{(3)} - \alpha_5^{(3)}} \H_{3,1}^{(3)}
  + \p{2\alpha_{3,2}^{(3)} - \alpha_{4,3}^{(3)} - \alpha_5^{(3)}}
  \H_{3,2}^{(3)}.
  \label{eq:H_3_3}
\end{align}
Note that the aforementioned divergence and its cancellation are
buried in $\alpha_{3,2}^{(3)}$.  Formally, this factor is calculated
by imposing an ultraviolet cutoff $\Lambda$ for the maximum values of
motional state indices $\ell,m,n$, and then taking the limit
$\Lambda\to\infty$.  This procedure ensures that there are no
divergences in $\alpha_{3,2}^{(3)}$.

\subsection{Effective four-body interactions at third order}

At third order in the coupling constants, we have four-body terms of
the form
\begin{align}
  \begin{tikzpicture}
    \begin{feynman}
      \vertex (v1);
      \vertex[above right = 1.5em of v1] (v2);
      \vertex[below right = 2.5em of v1] (v3);
      \vertex[right = 0.5em of v1] (label)
      {$\let\scriptstyle\textstyle\substack{m\mu q'\\n\nu r'}$};
      \vertex[above left = of v1] (f1) {$\mu q$};
      \vertex[below left = of v1] (f2) {$\nu r$};
      \vertex[above left = of v2] (f3) {$\rho s$};
      \vertex[below left = of v3] (f4) {$\sigma t$};
      \vertex[above right = of v2] (f5) {$\rho s'$};
      \vertex[right = of v2] (f6) {$\mu q''$};
      \vertex[right = of v3] (f7) {$\nu r''$};
      \vertex[below right = of v3] (f8) {$\sigma t'$};
      \diagram* {
        (f1) -- (v1) --[scalar] (v2) -- (f6),
        (f2) -- (v1) --[scalar] (v3) -- (f7),
        (f3) -- (v2) -- (f5),
        (f4) -- (v3) -- (f8),
      };
    \end{feynman}
  \end{tikzpicture}
  \propto K_{mn} K_m K_n G^{qr}_{q'r'} G^{q's}_{q''s'} G^{r't}_{r''t'}
  \c_{\mu q''}^\dag \c_{\nu r''}^\dag \c_{\rho s'}^\dag \c_{\sigma t'}^\dag
  \c_{\sigma t} \c_{\rho s} \c_{\nu r} \c_{\mu q}
  \label{eq:H_4_3_B}
\end{align}
and its mirror image, as well as
\begin{align}
  \begin{tikzpicture}
    \begin{feynman}
      \vertex[label=0:$~m\mu q'$] (v1);
      \vertex[below right = of v1, label=0:$~n\mu'q''$] (v2);
      \vertex[below right = of v2] (v3);
      \vertex[above left = of v1] (f1) {$\mu q$};
      \vertex[below left = of v1] (f2) {$\nu r$};
      \vertex[below left = of v2] (f3) {$\rho s$};
      \vertex[below left = of v3] (f4) {$\sigma t$};
      \vertex[above right = of v1] (f5) {$\nu r'$};
      \vertex[above right = of v2] (f6) {$\rho's'$};
      \vertex[above right = of v3] (f7) {$\sigma t'$};
      \vertex[below right = of v3] (f8) {$\mu'q'''$};
      \diagram* {
        (f1) -- (v1) -- (f5),
        (f2) -- (v1) -- (f5),
        (v1) --[scalar] (v2) --[scalar] (v3) -- (f8),
        (f3) -- (v2) -- (f6),
        (f4) -- (v3) -- (f7),
      };
    \end{feynman}
  \end{tikzpicture}
  \propto K_m K^m_n K_n
  G^{qr}_{q'r'} G^{\mu q'\rho s}_{\mu'q''\rho's'} G^{q''t}_{q'''t'}
  \c_{\mu'q'''}^\dag \c_{\nu r'}^\dag \c_{\rho's'}^\dag \c_{\sigma t'}^\dag
  \c_{\sigma t} \c_{\rho s} \c_{\nu r} \c_{\mu q}.
  \label{eq:H_4_3_C}
\end{align}
As we are computing the leading-order contribution to effective
four-body interactions, there are no counter-terms contributions.  In
principle, there is now also the possibility to make the disconnected
diagrams of the form
\begin{align}
  \shrink{
    \begin{tikzpicture}
      \begin{feynman}
        \vertex (v1);
        \vertex[above left = of v1] (f1);
        \vertex[below left = of v1] (f2);
        \vertex[right = of v1] (v2);
        \vertex[above right = of v2] (f3);
        \vertex[below right = of v2] (f4);
        \vertex[below = 1em of f2] (f5);
        \vertex[below = of f5] (f6);
        \vertex[below = 1em of f4] (f7);
        \vertex[below = of f7] (f8);
        \diagram* {
          (f1) -- (v1),
          (f2) -- (v1),
          (v2) -- (f3),
          (v2) -- (f4),
          (v1) --[scalar, half left] (v2),
          (v1) --[scalar, half right] (v2),
          (f5) -- (f8),
          (f6) -- (f7) };
      \end{feynman}
    \end{tikzpicture}},
  &&
  \shrink{
    \begin{tikzpicture}
      \begin{feynman}
        \vertex (v1);
        \vertex[above left = of v1] (f1);
        \vertex[below left = of v1] (f2);
        \vertex[right = of v1] (v2);
        \vertex[above right = of v2] (f3);
        \vertex[below right = of v2] (f4);
        \vertex[left = of f2] (a1);
        \vertex[left = of f4] (a2);
        \vertex[below = 1em of a1] (f5);
        \vertex[below = of f5] (f6);
        \vertex[below = 1em of a2] (f7);
        \vertex[below = of f7] (f8);
        \diagram* {
          (f1) -- (v1),
          (f2) -- (v1),
          (v2) -- (f3),
          (v2) -- (f4),
          (v1) --[scalar, half left] (v2),
          (v1) --[scalar, half right] (v2),
          (f5) -- (f8),
          (f6) -- (f7) };
      \end{feynman}
    \end{tikzpicture}},
  &&
  \t{and}
  &&
  \shrink{
    \begin{tikzpicture}
      \begin{feynman}
        \vertex (v1);
        \vertex[above left = of v1] (f1);
        \vertex[below left = of v1] (f2);
        \vertex[right = of v1] (v2);
        \vertex[above right = of v2] (f3);
        \vertex[below right = of v2] (f4);
        \vertex[right = of f2] (a1);
        \vertex[right = of f4] (a2);
        \vertex[below = 1em of a1] (f5);
        \vertex[below = of f5] (f6);
        \vertex[below = 1em of a2] (f7);
        \vertex[below = of f7] (f8);
        \diagram* {
          (f1) -- (v1),
          (f2) -- (v1),
          (v2) -- (f3),
          (v2) -- (f4),
          (v1) --[scalar, half left] (v2),
          (v1) --[scalar, half right] (v2),
          (f5) -- (f8),
          (f6) -- (f7) };
      \end{feynman}
    \end{tikzpicture}}.
\end{align}
As prescribed by $H_{\t{int}}^{(3)}$ in \eqref{eq:H_int_3}, however,
the second and third of these diagrams pick up a factor of $-1/2$, so
the sum over disconnected diagrams vanishes.

The contribution to the third-order four-body interaction Hamiltonian
from diagrams of the form in \eqref{eq:H_4_3_B} is
\begin{align}
  \shrink{
    \begin{tikzpicture}
      \begin{feynman}
        \vertex (v1);
        \vertex[above right = 1.5em of v1] (v2);
        \vertex[below right = 2.5em of v1] (v3);
        \vertex[above left = of v1] (f1);
        \vertex[below left = of v1] (f2);
        \vertex[above left = of v2] (f3);
        \vertex[below left = of v3] (f4);
        \vertex[above right = of v2] (f5);
        \vertex[right = of v2] (f6);
        \vertex[right = of v3] (f7);
        \vertex[below right = of v3] (f8);
        \diagram* {
          (f1) -- (v1) --[scalar] (v2) -- (f6),
          (f2) -- (v1) --[scalar] (v3) -- (f7),
          (f3) -- (v2) -- (f5),
          (f4) -- (v3) -- (f8),
        };
      \end{feynman}
    \end{tikzpicture}}
  - \f12 \shrink{
    \begin{tikzpicture}
      \begin{feynman}
        \vertex (v1);
        \vertex[above right = 1.5em of v1] (v2);
        \vertex[below right = 2.5em of v1] (v3);
        \vertex[above left = of v1] (f1);
        \vertex[below left = of v1] (f2);
        \vertex[above left = of v2] (f3);
        \vertex[below left = of v3] (f4);
        \vertex[above right = of v2] (f5);
        \vertex[right = of v2] (f6);
        \vertex[right = of v3] (f7);
        \vertex[below right = of v3] (f8);
        \diagram* {
          (f1) -- (v1) --[scalar] (v2) -- (f6),
          (f2) -- (v1) -- (v3) -- (f7),
          (f3) -- (v2) -- (f5),
          (f4) -- (v3) -- (f8),
        };
      \end{feynman}
    \end{tikzpicture}}
  = \p{\alpha_{4,1}^{(3)} - \f12\alpha_5^{(3)}} \H_{4,1},
\end{align}
where
\begin{align}
  \alpha_{4,1}^{(3)}
  \equiv \sum_{\substack{m\ge0\\n>0}} \f{K_{mn} K_m K_n}{E_{mn} E_n},
\end{align}
and
\begin{align}
  \H_{4,1}^{(3)}
  \equiv \sum_{\abs{\set{\mu,\nu,\rho,\sigma}}=4}
  G^{qr}_{q'r'} G^{q's}_{q''s'} G^{r't}_{r''t'}
  \c_{\mu q''}^\dag \c_{\nu r''}^\dag \c_{\rho s'}^\dag \c_{\sigma t'}^\dag
  \c_{\sigma t} \c_{\rho s} \c_{\nu r} \c_{\mu q}.
\end{align}
An equal contribution comes from the mirror images of these diagrams.
The contribution from diagrams of the form in \eqref{eq:H_4_3_C},
meanwhile, is
\begin{align}
  \shrink{
    \begin{tikzpicture}
      \begin{feynman}
        \vertex (v1);
        \vertex[below right = of v1] (v2);
        \vertex[below right = of v2] (v3);
        \vertex[above left = of v1] (f1);
        \vertex[below left = of v1] (f2);
        \vertex[below left = of v2] (f3);
        \vertex[below left = of v3] (f4);
        \vertex[above right = of v1] (f5);
        \vertex[above right = of v2] (f6);
        \vertex[above right = of v3] (f7);
        \vertex[below right = of v3] (f8);
        \diagram* {
          (f1) -- (v1) -- (f5),
          (f2) -- (v1) -- (f5),
          (v1) --[scalar] (v2) --[scalar] (v3) -- (f8),
          (f3) -- (v2) -- (f6),
          (f4) -- (v3) -- (f7),
        };
      \end{feynman}
    \end{tikzpicture}}
  - \f12 \shrink{
    \begin{tikzpicture}
      \begin{feynman}
        \vertex (v1);
        \vertex[below right = of v1] (v2);
        \vertex[below right = of v2] (v3);
        \vertex[above left = of v1] (f1);
        \vertex[below left = of v1] (f2);
        \vertex[below left = of v2] (f3);
        \vertex[below left = of v3] (f4);
        \vertex[above right = of v1] (f5);
        \vertex[above right = of v2] (f6);
        \vertex[above right = of v3] (f7);
        \vertex[below right = of v3] (f8);
        \diagram* {
          (f1) -- (v1) -- (f5),
          (f2) -- (v1) -- (f5),
          (v1) --[scalar] (v2) -- (v3) -- (f8),
          (f3) -- (v2) -- (f6),
          (f4) -- (v3) -- (f7),
        };
      \end{feynman}
    \end{tikzpicture}}
  - \f12 \shrink{
    \begin{tikzpicture}
      \begin{feynman}
        \vertex (v1);
        \vertex[below right = of v1] (v2);
        \vertex[below right = of v2] (v3);
        \vertex[above left = of v1] (f1);
        \vertex[below left = of v1] (f2);
        \vertex[below left = of v2] (f3);
        \vertex[below left = of v3] (f4);
        \vertex[above right = of v1] (f5);
        \vertex[above right = of v2] (f6);
        \vertex[above right = of v3] (f7);
        \vertex[below right = of v3] (f8);
        \diagram* {
          (f1) -- (v1) -- (f5),
          (f2) -- (v1) -- (f5),
          (v1) -- (v2) --[scalar] (v3) -- (f8),
          (f3) -- (v2) -- (f6),
          (f4) -- (v3) -- (f7),
        };
      \end{feynman}
    \end{tikzpicture}}
  = \p{\alpha_{4,2}^{(3)} - \alpha_5^{(3)}} \H_{4,2},
\end{align}
where
\begin{align}
  \alpha_{4,2}^{(3)}
  \equiv \sum_{m,n>0} \f{K_m K^m_n K_n}{E_m E_n},
\end{align}
and
\begin{align}
  \H_{4,2}^{(3)}
  \equiv \sum_{\abs{\set{\mu,\nu,\rho,\sigma}}=4}
  G^{qr}_{q'r'} G^{\mu q'\rho s}_{\mu'q''\rho's'} G^{q''t}_{q'''t'}
  \c_{\mu'q'''}^\dag \c_{\nu r'}^\dag \c_{\rho's'}^\dag \c_{\sigma t'}^\dag
  \c_{\sigma t} \c_{\rho s} \c_{\nu r} \c_{\mu q}.
\end{align}
The net third-order four-body interaction Hamiltonian is therefore
\begin{align}
  H_4^{(3)}
  = \p{2\alpha_{4,1}^{(3)} - \alpha_5^{(3)}} \H_{4,1}^{(3)}
  + \p{\alpha_{4,2}^{(3)} - \alpha_5^{(3)}} \H_{4,2}^{(3)}.
  \label{eq:H_4_3}
\end{align}

\section{Low-excitation Hamiltonians, eigenstates, and spectra}
\label{sec:spectra}

Current experiments with ultracold ${}^{87}$Sr on a lattice can
coherently address ground states and single orbital excitations of up
to five atoms per lattice site \cite{goban2018emergence}.  Due to the
SU($N$) symmetry of inter-atomic interactions, manifest in the fact
that all coupling constants are independent of nuclear spin, a
restriction of the $M$-body Hamiltonians $H_M=\sum_p H_M^{(p)}$ to the
subspace of experimentally addressed states takes the form
\begin{align}
  H_M = \sum_{\abs{\set{\mu_j}}=M}
  \p{U_{M,\g} \n_{\mu_1,\g} \n_{\mu_2,\g}
    + U_{M,+} \n_{\mu_1,\e} \n_{\mu_2,\g}
    + U_{M,-} \c_{\mu_1,\g}^\dag \c_{\mu_2,\e}^\dag
    \c_{\mu_2,\g} \c_{\mu_1,\e}}
  \prod_{\alpha=3}^M \n_{\mu_\alpha,\g},
  \label{eq:H_M}
\end{align}
where $\n_{\mu s}\equiv \c_{\mu s}^\dag\c_{\mu s}$ is a number
operator, and the coefficients $U_X$ can be determined from the
coupling constants $G_Y$ and prefactors $\alpha_Z^{(p)}$ of the
effective $M$-body Hamiltonians derived in section
\ref{sec:low_energy} (see Appendix \ref{sec:U_X}).  For a lattice with
$N\ge M$ atoms occupying nuclear spins $\N=\set{\mu_j}$ for
$j=1,2,\cdots,N$, the $M$-body Hamiltonian $H_M$ has a single ground
state $\ket{\N,0}$, and a singly-excited state $\ket{\N,+}$ which is
fully symmetric in the orbital degrees of freedom; these states are
\begin{align}
  \ket{\N,0}
  \equiv \p{\prod_{\mu\in\N} \c_{\mu,\g}^\dag} \ket{\t{vacuum}},
  &&
  \ket{\N,+} \equiv \f1{\sqrt{N}} \sum_{\mu\in\N}
  \c_{\mu,\e}^\dag \c_{\mu,\g} \ket{\N,0}.
  \label{eq:states_S}
\end{align}
As these states are fully symmetric in their orbital degrees of
freedom, they are anti-symmetric in their nuclear spin degrees of
freedom, forming an SU($N$) singlet.  Furthermore, the symmetric state
is particularly interesting as its orbital degrees of freedom form an
$N$-body entangled $W$ state, which belongs to a special class of
multi-partite entangled states that are robust against disposal or
loss of particles.  This state thus constitutes an important resource
for many quantum information processing and quantum communication
tasks \cite{zang2015generating}.

In addition to the states in \eqref{eq:states_S}, the multi-body
Hamiltonian $H_M$ in \eqref{eq:H_M} has an $\p{N-1}$-fold degenerate
excited-state eigenspace which is asymmetric in the orbital degrees of
freedom, spanned by the states
\begin{align}
  \ket{\N,-,j} \equiv \f1{\sqrt2}
  \p{\c_{\mu_1,\e}^\dag \c_{\mu_1,\g} - \c_{\mu_j,\e}^\dag \c_{\mu_j,\g}}
  \ket{\N,0}
  \label{eq:states_A}
\end{align}
for $j=2,\cdots,N$.  If $N>2$, the asymmetric states are not separable
in their orbital and nuclear spin degrees of freedom.  An important
feature of the excited states in \eqref{eq:states_S} and
\eqref{eq:states_A} is that they are entirely independent of $M$,
which implies that the effect of multi-body interactions is simply to
modify the many-body atomic energy spectra without affecting the
energy eigenstates.  The eigenvalues $E_{NX}^{(M)}=\bk{\N X|H_M|\N X}$
of $H_M$ associated with each of the eigenstates in
\eqref{eq:states_S} and \eqref{eq:states_A} are provided in table
\ref{tab:eigen}, both in terms of the coefficients $U_{MX}$ of $H_M$
as expressed in \eqref{eq:H_M} and the $M$-body eigenvalues
$E_{MX}^{(M)}$.  Due to the SU($N$) symmetry of the multi-body
Hamiltonian $H_M$, the eigenvalues $E_{NX}^{(M)}$ depend on the number
of nuclear spins on a lattice site, $N$, but not on the actual nuclear
spins $\mu\in\N$ which are occupied.  The total $N$-body interaction
energies $E_{NX}$ are given in terms of the $M$-body eigenvalues
$E_{NX}^{(M)}$ by $E_{NX}=\sum_ME_{NX}^{(M)}$.

\begin{table}
  \centering
  \caption{\footnotesize Low-excitation eigenvalues of $M$-body
    Hamiltonians $H_M$.  Many-body energy eigenstates are labeled by
    the nuclear spins they occupy (i.e.~$\N$ with $N\equiv\abs{\N}$)
    and whether they are in an orbital ground ($0$), singly-excited
    symmetric ($+$), or singly-excited asymmetric ($-$) state.  The
    corresponding $N$-body eigenvalues $E_{NX}^{(M)}$ of $H_M$ are
    given in terms of the coefficients $U_{MX}$ as appearing in
    \eqref{eq:H_M} (first three rows), in addition to the $M$-body
    eigenvalues $E_{MX}^{(M)}$ (last three rows).}
  \label{tab:eigen}
  \begin{tabular}{c|c}
    Eigenstate
    & $H_M$ eigenvalue ($M\le N$) \\ \hline\hline
    $\ket{\N,0}$
    & $M! {N \choose M} U_{M,\g}$ \\ \hline
    $\ket{\N,+}$
    & $M! {N-1 \choose M} U_{M,\g}
    + \p{M-1}! {N-1 \choose M-1} \p{U_{M,+} + U_{M,-}}$ \\ \hline
    $\ket{\N,-}$
    & $M! {N-1 \choose M} U_{M,\g}
    + \p{M-1}! {N-1 \choose M-1} U_{M,+}
    - \p{M-2}! {N-2 \choose M-2} U_{M,-}$ \\ \hline\hline
    $\ket{\N,0}$
    & ${N \choose M} E_{M,0}^{(M)}$ \\ \hline
    $\ket{\N,+}$
    & ${N-1 \choose M} E_{M,0}^{(M)}
    + {N-1 \choose M-1} E_{M,+}^{(M)}$ \\ \hline
    $\ket{\N,-}$
    & ${N-1 \choose M} E_{M,0}^{(M)}
    + M^{-1} \sp{{N-1 \choose M-1} - {N-2 \choose M-2}} E_{M,+}^{(M)}
    + \sp{\p{1 - M^{-1}} {N-1 \choose M-1}
      + M^{-1} {N-2 \choose M-2}} E_{M,-}^{(M)}$
  \end{tabular}
\end{table}

\subsection{Many-body state spectroscopy}

Spectroscopic interrogation is a powerful means to probe the internal
structure and dynamics of a system under examination.  Consequently,
we consider Rabi spectroscopy of the low-lying energy eigenstates in
multiply-occupied lattice sites.  If we interrogate a lattice site by
a laser red-detuned by $\Delta$ from the single-atom orbital state
excitation energy, we realize the Hamiltonian
\begin{align}
  H_{\t{Rabi}}
  = \sum_X E_X \P_X
  + \sum_\mu\p{\Delta T_\mu^\z + \Omega_\mu T_\mu^\x},
  \label{eq:H_rabi}
\end{align}
where $E_X$ is an eigenvalue of the effective interaction Hamiltonian
$H_{\t{int}}^{\t{eff}}$, $\P_X$ is a projector onto the corresponding
eigenspace, and
\begin{align}
  T_\mu^\z \equiv \f12 \p{\c_{\mu,\e}^\dag \c_{\mu,\e}
    - \c_{\mu,\g}^\dag \c_{\mu,\g}},
  &&
  T_\mu^\x \equiv \f12 \p{\c_{\mu,\e}^\dag \c_{\mu,\g}
    + \c_{\mu,\g}^\dag \c_{\mu,\e}},
\end{align}
are single-atom pseudospin operators.  The Rabi frequency $\Omega_\mu$
is proportional to the Clebsch-Gordan coefficient
$\bk{I,\mu;1,0|I,\mu}\propto\mu$ for a photon-induced
nuclear-spin-conserving orbital state transition of an atom with
nuclear spin $\mu$.  We therefore define the ``bare'' Rabi frequency
$\Omega_0\equiv\Omega_\mu/\mu$ to explicitly factor out dependence on
nuclear spins $\mu$.

Consider now a single lattice site in the orbital ground state
$\ket{\N,0}$ with nuclear spins $\N\equiv\set{\mu_j}$ for
$j=1,2,\cdots,N$.  If we red-detune the interrogation laser by
$\delta$ from a many-body orbital state excitation energy, i.e.~set
$\Delta=\Delta_{NX}-\delta$ for $\Delta_{NX}\equiv E_{NX}-E_{N,0}$ and
$X\in\set{+,-}$, then in the subspace of the target states
$\set{\ket{\N,0},\ket{\N X}}$ the Hamiltonian in \eqref{eq:H_rabi}
becomes
\begin{align}
  H_{\N X} = \delta S_{\N X}^\z + \Omega_{\N X} S_{\N X}^\x,
  \label{eq:H_NX}
\end{align}
where
\begin{align}
  S_{\N X}^\z &\equiv \f12\p{\op{\N X}-\op{\N,0}},
  \label{eq:S_NX_z} \\
  S_{\N X}^\x &\equiv \f12\p{\op{\N X}{\N,0}+\op{\N,0}{\N X}},
  \label{eq:S_NX_x}
\end{align}
are many-body pseudospin operators, and the Rabi frequencies
$\Omega_{\N X}$ are determined by
\begin{align}
  H_{\t{Rabi}}\ket{\N,0}
  = \f12\Omega_0\sum_\mu\mu\c_{\mu,\e}^\dag\c_{\mu,\g}\ket{\N,0}
  = \f12\Omega_{\N,+}\ket{\N,+} + \f12\Omega_{\N,-}\ket{\N,-}.
  \label{eq:O_NX}
\end{align}
While the symmetric excited state $\ket{\N,+}$ is given in
\eqref{eq:states_S}, at this point we have not explicitly solved for
the asymmetric excited state $\ket{\N,-}$.  The asymmetric state is
implicitly defined by \eqref{eq:O_NX}, and lies somewhere in the span
of the $N-1$ asymmetric states given in \eqref{eq:states_A}.
Determining the symmetric-state Rabi frequency $\Omega_{\N,+}$ is
simply a matter of projecting the expression in \eqref{eq:O_NX} onto
$\ket{\N,+}$, which yields
\begin{align}
  \Omega_{\N,+}
  = \bk{\N,+|\Omega_0\sum_\mu\mu\c_{\mu,\e}^\dag\c_{\mu,\g}|\N,0}
  = \Omega_0 \sum_{\mu\in\N} \f{\mu}{\sqrt{N}}
  = \Omega_0 \sqrt{N} \bar\mu_\N,
  \label{eq:O_N+}
\end{align}
where $\bar\mu_\N\equiv\sum_{\mu\in\N}\mu/N$ is the average nuclear
spin $\mu\in\N$.  In order to determine the asymmetric-state Rabi
frequency $\Omega_{\N,-}$, we rearrange \eqref{eq:O_NX} to find
\begin{align}
  \Omega_{\N,-}\ket{\N,-}
  = \Omega_0\sum_\mu\mu\c_{\mu,\e}^\dag\c_{\mu,\g}\ket{\N,0}
  - \Omega_{\N,+}\ket{\N,+}
  = \Omega_0 \sum_{\mu\in\N} \p{\mu - \bar\mu_\N}
  \c_{\mu,\e}^\dag \c_{\mu,\g} \ket{\N,0}.
\end{align}
Denoting the standard deviation of nuclear spins $\mu\in\N$ by
$\sigma_\N$, normalization of $\ket{\N,-}$ thus determines the
asymmetric-state Rabi frequency
\begin{align}
  \Omega_{\N,-}
  = \Omega_0 \sp{\sum_{\mu\in\N}\p{\mu-\bar\mu_\N}^2}^{1/2}
  = \Omega_0 \sqrt{N} \sigma_\N,
  \label{eq:O_N-}
\end{align}
which in turn implies that the asymmetric excited state $\ket{\N,-}$
is
\begin{align}
  \ket{\N,-} = \f1{\sqrt{N}}\sum_{\mu\in\N}
  \p{\f{\mu-\bar\mu_\N}{\sigma_\N}}
  \c_{\mu,\e}^\dag \c_{\mu,\g} \ket{\N,0}.
\end{align}

\begin{figure}
  \centering
  \includegraphics{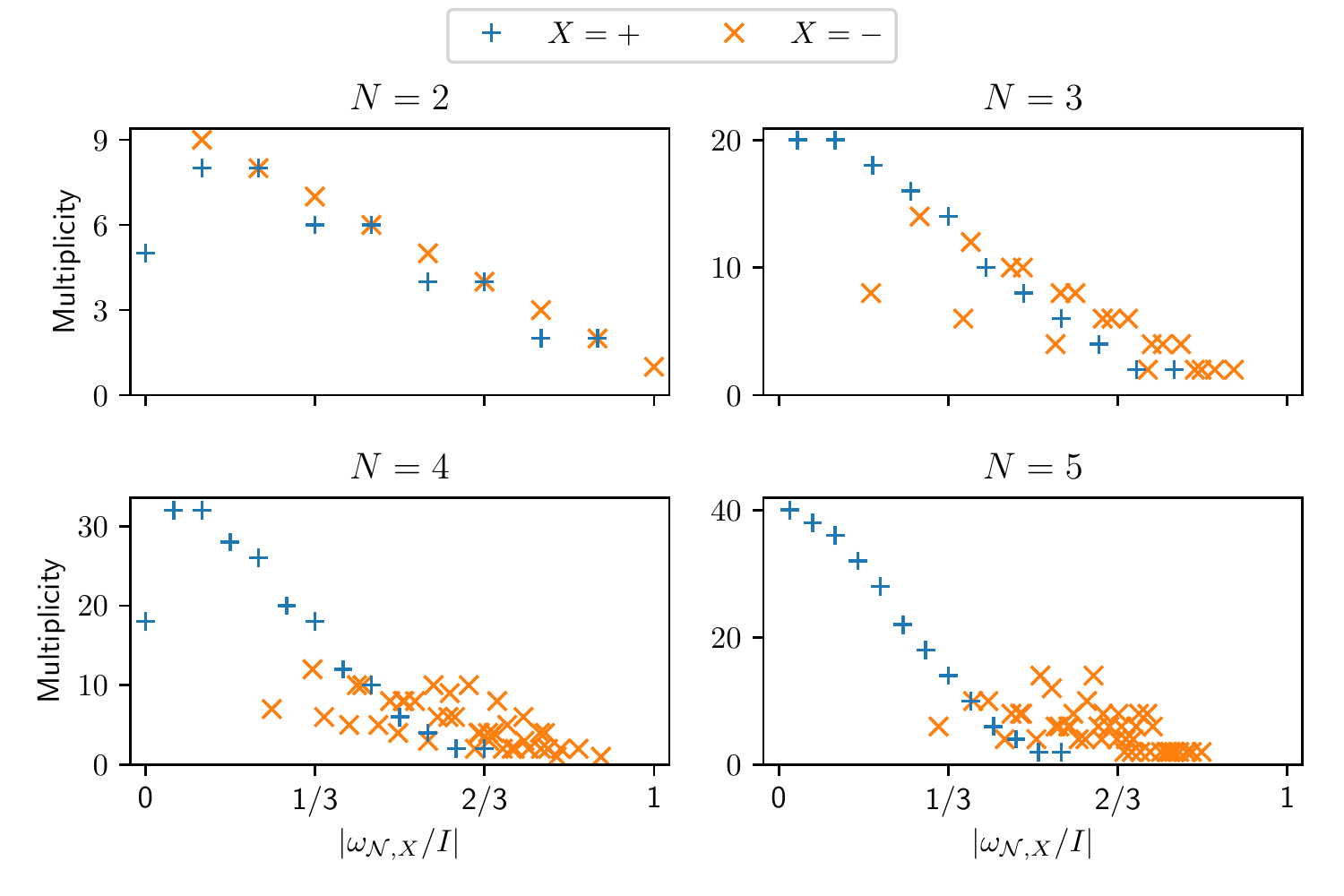}
  \caption{\footnotesize Multiplicities of the magnitudes of reduced
    Rabi frequencies
    $\omega_{\N X}\equiv\Omega_{\N X}/\Omega_0\sqrt{N}$ in a lattice
    with a uniform mixture of nuclear spins with $I=9/2$ and
    single-site occupation numbers $N$ which are achievable in current
    ${}^{87}$Sr experiments.}
  \label{fig:coefficients}
\end{figure}

Figure \ref{fig:coefficients} shows multiplicities of the magnitudes
of reduced Rabi frequencies
$\omega_{\N X}\equiv\Omega_{\N X}/\Omega_0\sqrt{N}$ in a lattice with
a uniform mixture of nuclear spins with $I=9/2$ for single-site
occupation numbers $N$ which are achievable in current ${}^{87}$Sr
experiments \cite{goban2018emergence}.  On average, asymmetric-state
Rabi frequencies are greater in magnitude, which becomes more
pronounced for larger single-site occupation numbers.

\subsection{Experimental signatures and comparison}

\begin{figure}
  \centering
  \subfloat{\includegraphics{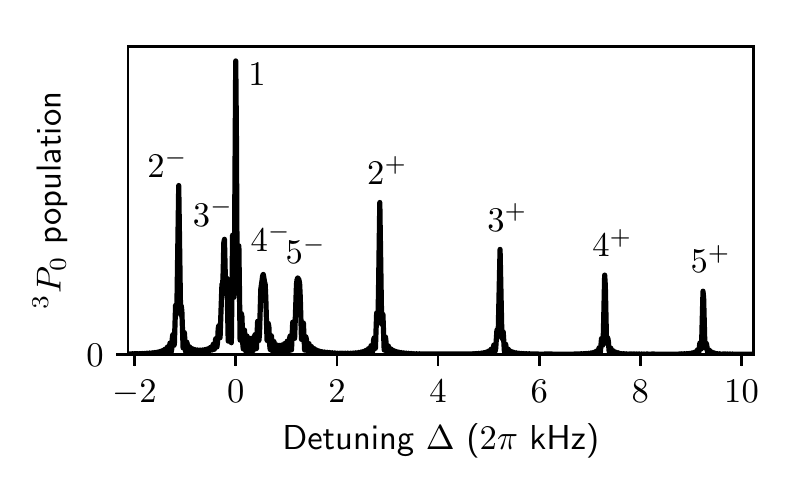}}
  \subfloat{\includegraphics{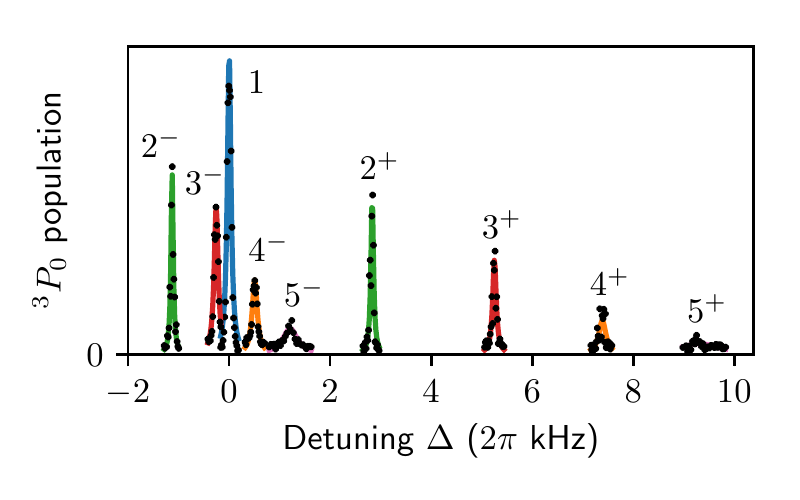}}
  \caption{\footnotesize Population (in arbitrary units) of the
    excited ${}^3P_0$ orbital state of ${}^{87}$Sr atoms in a uniform
    mixture of nuclear spins.  Atoms are prepared in the ground state
    of a lattice with depth $\U=54 E_{\t{R}}$, where
    $E_{\t{R}}\approx3.5\times2\pi~\t{kHz}$ is the lattice photon
    recoil energy of the atoms, and interrogated by a laser with Rabi
    frequency $\Omega_I=50\times2\pi~\t{Hz}$ for a time
    $t=\pi/\Omega_I$.  ({\bf Left}) Populations predicted by the
    low-energy effective theory (with $s$-wave scattering parameters
    retrieved from ref.~\cite{zhang2014spectroscopic}), averaged over
    all nuclear spin combinations of $N\in\set{1,\cdots,5}$ atoms per
    lattice site for a fixed total atom number.  ({\bf Right})
    Experimental measurements of ${}^3P_0$ populations retrieved from
    ref.~\cite{goban2018emergence}, with Lorentzian fits to each peak
    as a visual guide.  Resonance peaks are identified by the
    many-body orbital states which are excited at the
    peak.}
  \label{fig:sweep}
\end{figure}

We now consider samples of ${}^{87}$Sr atoms in a uniform mixture of
nuclear spins $\mu\in\set{-I,-I+1,\cdots,I}$ prepared in motional
ground states of a rectangular lattice with depths
$\U=\p{\U_\x,\U_\y,\U_\z}=\p{41,55,69} E_{\t{R}}$, where
$E_{\t{R}}\approx3.5\times2\pi~\t{kHz}$ is the lattice photon recoil
energy of the atoms.  Such samples can be prepared in experiments
which can vary the single-site occupation number $N$, and which can
control for the total number of atoms that are addressed by an
external interrogation laser.  Figure \ref{fig:sweep} shows the
population of the excited ${}^3P_0$ orbital state when these atoms are
interrogated for a time $t=\pi/\Omega_I$ by a laser with Rabi
frequency $\Omega_I=50\times2\pi~\t{Hz}$ (i.e.~for individual atoms
with nuclear spin $\mu=I$) and detuning $\Delta$ from the single-atom
${}^1S_0\to{}^3P_0$ orbital excitation energy.  The ${}^3P_0$
population peaks when the laser detuning $\Delta$ is equal to the
many-body excitation energy $\Delta_{NX}\equiv E_{NX}-E_{N,0}$ for
$X\in\set{+,-}$, as this is precisely when the on-resonance condition
$\delta=0$ is satisfied in the many-body Rabi Hamiltonian $H_{\N X}$
in \eqref{eq:H_NX}.  Due to experimental uncertainties which vary with
single-site occupation number $N$, the heights of experimental peaks
in figure \ref{fig:sweep} are not well-calibrated between different
values of $N$.  Nonetheless, figure \ref{fig:sweep} exhibits
signatures of larger asymmetric-state Rabi frequencies than
symmetric-state ones in the form of higher asymmetric-state peaks for
fixed $N$.

\begin{figure}
  \centering
  \includegraphics{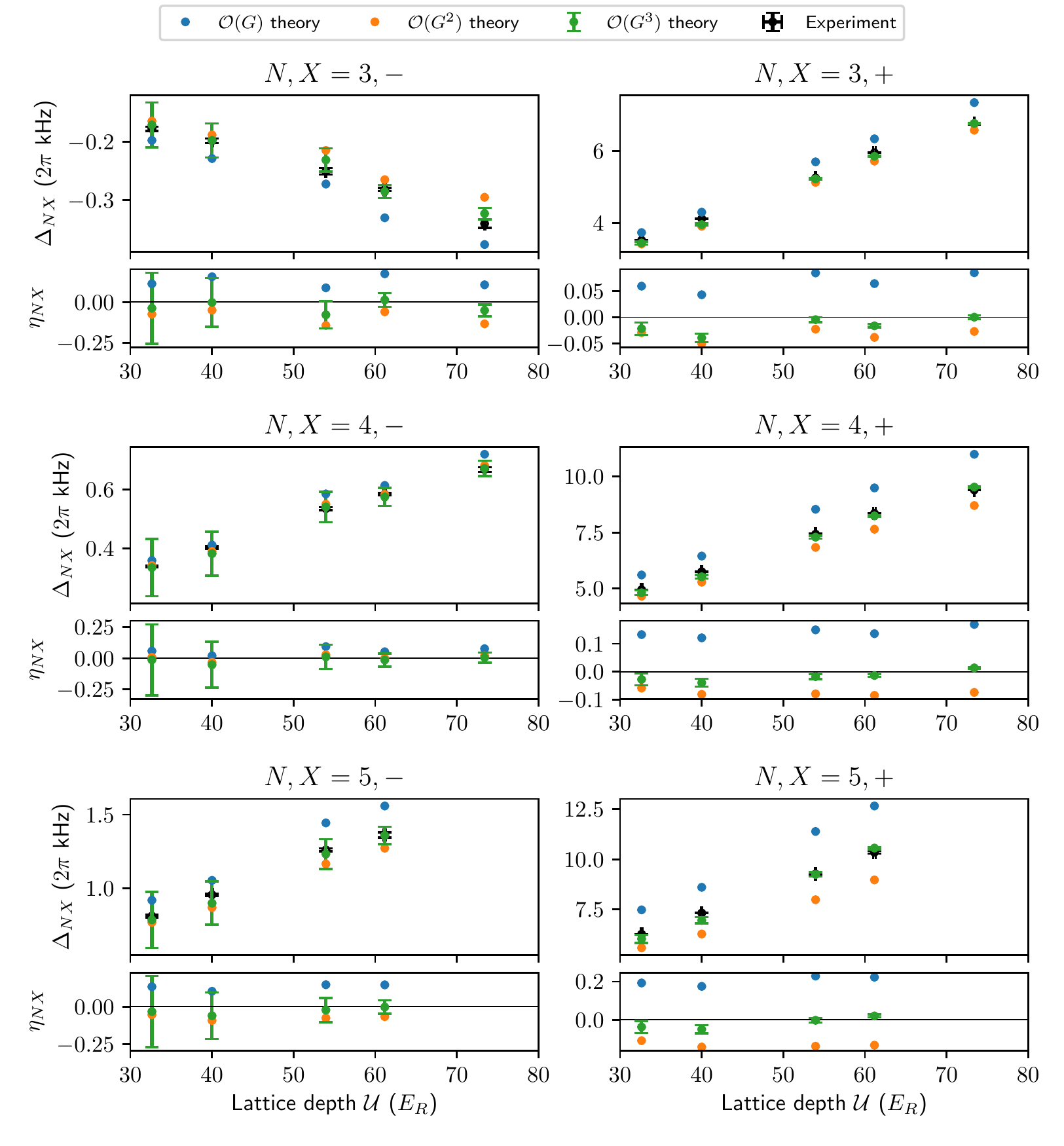}
  \caption{\footnotesize Multi-body excitation energies of ultracold
    ${}^{87}$Sr atoms at various lattice depths.  The top plot in each
    sub-figure with fixed $N,X$ shows the excitation energies
    $\Delta_{NX}\equiv E_{NX}-E_{N,0}$ measured experimentally in
    ref.~\cite{goban2018emergence} and those predicted by the
    low-energy effective theory at different orders in the coupling
    constants, when applicable both with and without four-body
    contributions.  The bottom plot in each sub-figure shows the
    relative error
    $\eta_{NX}\equiv
    \Delta_{NX}^{\t{theory}}/\Delta_{NX}^{\t{experiment}}-1$.  Error
    bars represent experimental error or conservatively estimated
    theoretical uncertainties from nearest-neighbor hopping of virtual
    states in the low-energy effective theory (see Appendix
    \ref{sec:error}).}
  \label{fig:shifts}
\end{figure}

\begin{figure}
  \centering
  \includegraphics{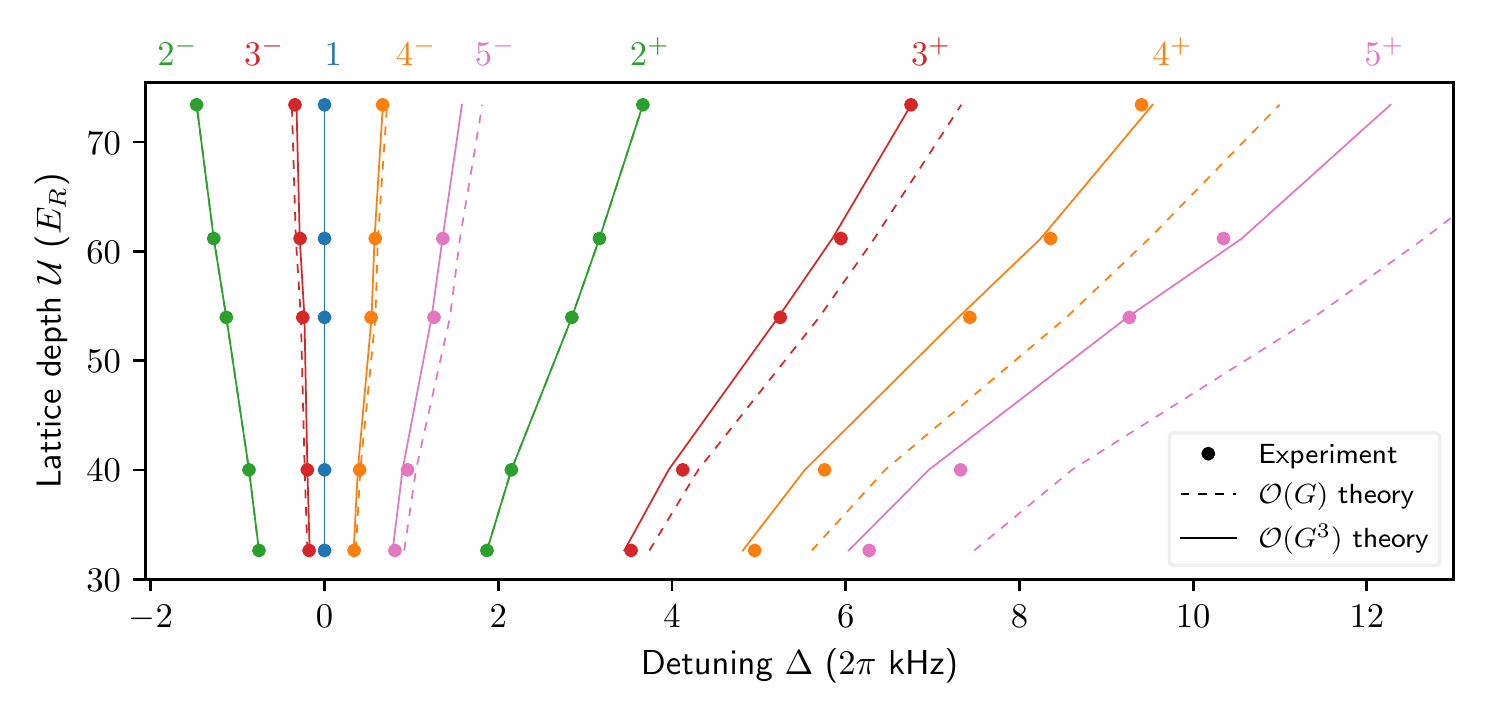}
  \caption{\footnotesize Summary of the many-body excitation spectra
    in figure \ref{fig:shifts}, retrieved from
    ref.~\cite{goban2018emergence}.}
  \label{fig:shifts_summary}
\end{figure}

Identifying peaks in excitation spectra such as in figure
\ref{fig:sweep} constitutes a measurement of many-body excitation
energies, which was performed in ref.~\cite{goban2018emergence} to
detect signatures of effective multi-body interactions.  Figure
\ref{fig:shifts} shows a comparison between
\begin{enumerate*}
\item experimental measurements of the many-body excitation energies
  $\Delta_{NX}$ for all $\p{N,X}\in\set{3,4,5}\times\set{+,-}$ at
  various mean lattice depths $\U$, and
\item the corresponding values of $\Delta_{NX}$ predicted by the
  low-energy effective theory at different orders in the coupling
  constants.
\end{enumerate*}
A known source of error in our effective theory comes from neglecting
the inter-site matrix elements of all Hamiltonians.  This error is
discussed in Appendix \ref{sec:error}, and leads to theoretical
uncertainties represented by error bars on the $\O\p{G^3}$ theory in
figure \ref{fig:shifts}.  A summary of figure \ref{fig:shifts} is
provided in figure \ref{fig:shifts_summary}.  We note that many-body
interaction energy shifts are smaller for asymmetric ($-$) states than
symmetric ($+$) ones due to the competition between contributions of
opposite sign in the asymmetric case (see rows 2 and 3 of table
\ref{tab:eigen}, where as a consequence of positive scattering lengths
in the case of $^{87}$Sr, all $U_{MX}$ for fixed $M$ have the same
sign).  This competition is a many-body analogue of the two-body case
with a competition between direct and exchange terms in the
interaction energies of singly-excited states.

The results in figures \ref{fig:shifts} and \ref{fig:shifts_summary}
highlight a few important points about ultracold, high-density
${}^{87}$Sr experiments and our low-energy effective theory.  First,
these experiments exhibit clear signatures of multi-body interactions,
as evidenced by a stark disagreement between the observed many-body
excitation energies $\Delta_{NX}$ and those that are predicted by the
two-body $\O\p{G}$ theory.  Multi-body interactions are thus crucial
for understanding these high-density experiments in the context of a
single-band Hubbard model, which naturally arises in the
zero-temperature limit when all atoms occupy their motional ground
state.  Second, the inter-atomic interactions in these experiments are
strong enough to require going beyond leading order for the
description of multi-body interactions in the low-energy effective
theory.  The formally small quantities organizing our perturbation
theory are two-body ground-state interaction energies (proportional to
the couplings $G_X$) divided by the spectral gap of the single-atom
Hamiltonian $H_0$.  These reduced (dimensionless) interaction energies
vary from $\sim0.05-0.15$ in the parameter regimes of the ${}^{87}$Sr
experiments considered here (see Appendix \ref{sec:pert_params}).  As
experiments begin to operate at higher atom densities with amplified
interaction effects, reliably predicting interaction energies may
require going to yet higher orders in perturbation theory.  Due to a
combinatorial explosion of the number of diagrams which appear at
increasing orders in the effective theory, however, we need more
systematic methods to compute effective multi-body Hamiltonians at
fourth order.  In any case, we are agnostic as to whether such a
calculation would provide better agreement between experiment and
theory without first performing a detailed analysis of systematic
errors.

\subsection{Orbital-state dynamics of a nuclear spin mixture}

In addition to spectral measurements of many-body interaction
energies, we consider the dynamics of multiply-occupied lattice sites
during spectroscopic interrogation.  While these dynamics do not
provide information about the nature or origin of effective multi-body
interactions, they provide tools and intuition for addressing the
low-lying orbital excitations which are readily accessible in an
experimental setting.  If we initialize all atoms in the $N$-body
ground state with an incoherent mixture of all nuclear spins, then we
prepare the mixed state $\rho_{N,0}=\P_{N,0}/\tr\P_{N,0}$, where
\begin{align}
  \P_{NX} \equiv \sum_{\abs{\N}=N}\op{\N X}
  \label{eq:P_NX}
\end{align}
is a projector onto the space of the $N$-body orbital states
$\ket{\N X}$.  Interrogating the atoms for a time $t$ by a laser
resonant with the excitation energy $\Delta_{N\pm}$ then gives us the
state
\begin{align}
  \rho_N^{(\pm)}\p{t} = \f1{\tr\P_{N,0}}
  \sum_{\abs{\N}=N} \exp\p{-it\Omega_{\N\pm}S_{\N\pm}^\x}
  \op{\N,0} \exp\p{it\Omega_{\N\pm}S_{\N\pm}^\x},
\end{align}
where the Rabi frequencies $\Omega_{\N,+},\Omega_{\N,-}$ and
pseudo-spin operators $S_{\N\pm}^\x$ are respectively given in
\eqref{eq:O_N+}, \eqref{eq:O_N-}, and \eqref{eq:S_NX_x}.  Denoting the
eigenstates of $S_{\N\pm}^\x$ by
\begin{align}
  \ket{\N,\S_\pm} \equiv \f1{\sqrt2} \p{\ket{\N,0} + \ket{\N\pm}},
  &&
  \ket{\N,\A_\pm} \equiv \f1{\sqrt2} \p{\ket{\N,0} - \ket{\N\pm}},
\end{align}
and defining the identity operator projected to the relevant subspace,
\begin{align}
  \1_{\N\pm} \equiv \op{\N,0} + \op{\N\pm}
  = \op{\N,\S_\pm} + \op{\N,\A_\pm},
\end{align}
we can write the state $\rho_N^{(\pm)}\p{t}$ and excited-state
projectors $\P_{N\pm}$ in the form
\begin{align}
  \rho_N^{(\pm)}\p{t} = \f1{\tr\P_{N,0}} \sum_{\abs{\N}=N} \f12
  \sp{\1_{\N\pm} + e^{i2t\Omega_{\N\pm}}\op{\N,\S_\pm}{\N,\A_\pm}
    + e^{-i2t\Omega_{\N\pm}}\op{\N,\A_\pm}{\N,\S_\pm}},
  \label{eq:rho_N_pm_t}
\end{align}
and
\begin{align}
  \P_{N\pm} = \sum_{\abs{\N}=N} \f12
  \sp{\1_{\N\pm}-\op{\N,\S_\pm}{\N,\A_\pm}-\op{\N,\A_\pm}{\N,\S_\pm}},
\end{align}
from which it follows that the net excited-state population at time
$t$ is
\begin{align}
  \bk{\P_{N\pm}\p{t}}
  \equiv \tr\sp{\rho_N^{(\pm)}\p{t}\P_{N\pm}}
  = \f12 - \f12 \bk{\cos\p{2t\Omega_{\N\pm}}}_{\abs{\N}=N},
  \label{eq:P_N_pm_t}
\end{align}
where $\bk{X}_{\abs{\N}=N}\equiv\sum_{\abs{\N}=N}X/\tr\P_{N,0}$ is an
average of $X$ over all choices of $N$ distinct nuclear spins.

\begin{figure}
  \centering
  \includegraphics{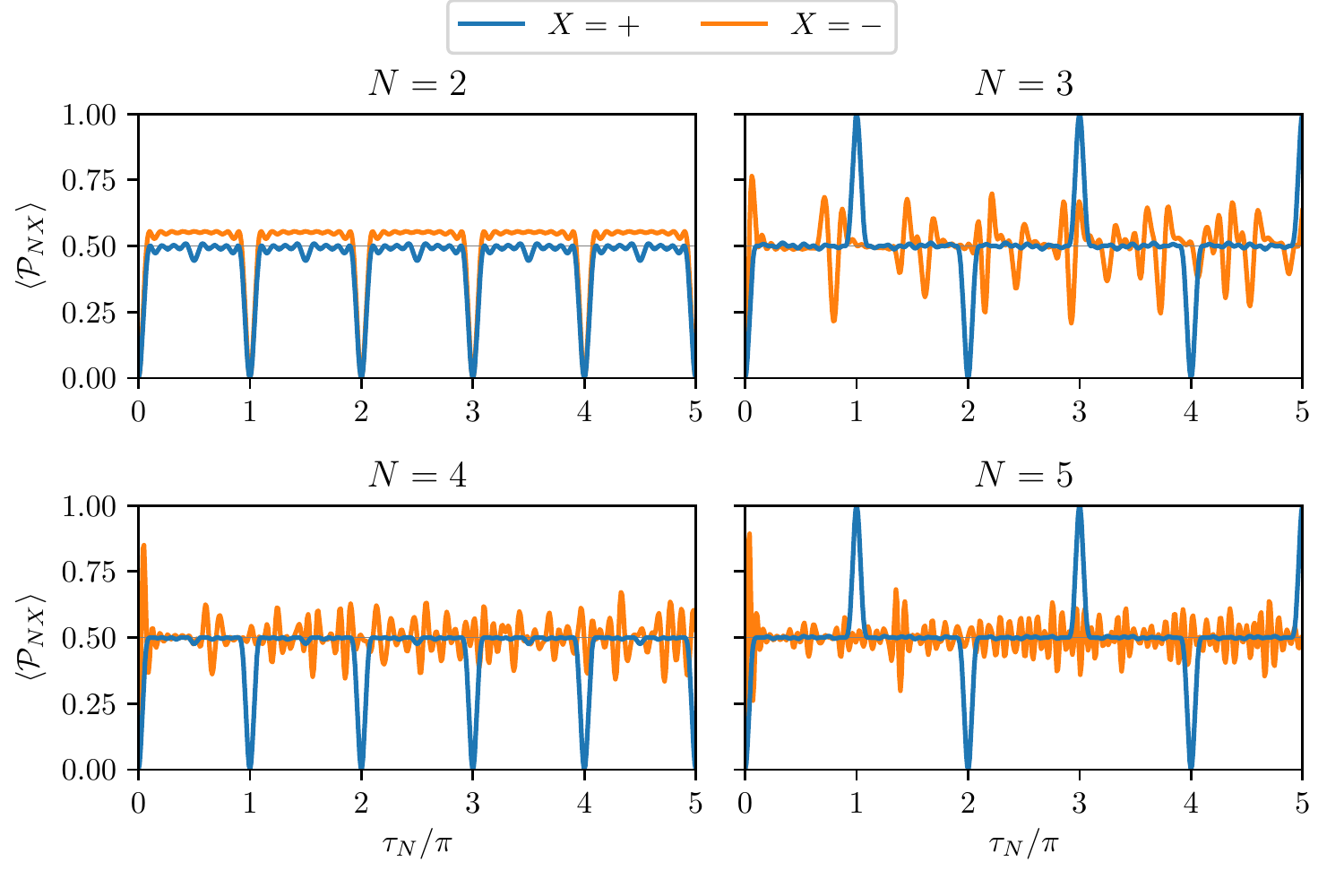}
  \caption{\footnotesize Net population of the $N$-body orbital
    excited states $\set{\ket{\N\pm}}$ after interrogation of an
    initial mixed state $\rho_{N,0}=\P_{N,0}/\tr\P_{N,0}$ for a
    reduced time $\tau_N\equiv t\Omega_0/\sqrt{N}$ (i.e.~with real
    time $t$) by a laser with bare Rabi frequency $\Omega_0$ which is
    resonant with the $N$-body excitation energy $\Delta_{NX}$.  Here
    $\P_{NX}$, defined in \eqref{eq:P_NX}, is a projector onto the
    space of the $N$-body orbital states $\ket{\N X}$.}
  \label{fig:time}
\end{figure}

Figure \ref{fig:time} shows the excited-state population
$\bk{\P_{N\pm}\p{t}}$ for several occupation numbers $N$.  With the
exception of $N=2$, the asymmetric-state populations generally have an
initial short period of growth before falling back to
$\bk{\P_{N,-}}\approx1/2$.  This behavior can be understood by the
fact that for fixed $N>2$, any pair of Rabi frequencies
$\Omega_{\N_1,-},\Omega_{\N_2,-}$ with
$\Omega_{\N_1,-}\ne\Omega_{\N_2,-}$ are mutually incommensurate, which
implies that at times $t$ with
$\min_{\abs{\N}=N}\set{2t\Omega_{\N,-}}\gtrsim1$ the averaging in
\eqref{eq:P_N_pm_t} effectively becomes a pseudo-random sampling
average of $\cos x$ over values of $x$, so
$\bk{\cos\p{2t\Omega_{\N,-}}}_{\abs{\N}=N}\approx0$.  When $N=2$, the
asymmetric-state Rabi frequencies essentially take on the same values
as the symmetric-state ones (see figure \ref{fig:coefficients}); the
behavior of asymmetric-state population dynamics for $N=2$ can
therefore be understood by the following discussion of symmetric-state
population dynamics.

To understand the periodic collapse and revival of symmetric-state
populations in figure \ref{fig:time}, we observe from \eqref{eq:O_N+}
that the symmetric-state phases in \eqref{eq:rho_N_pm_t} and
\eqref{eq:P_N_pm_t} take the form
\begin{align}
  2t\Omega_{\N,+} = \tau_N \sum_{\mu\in\N} 2\mu
  &&
  \t{with}
  &&
  \tau_N \equiv \f{t\Omega_0}{\sqrt{N}},
  \label{eq:2tO_N+}
\end{align}
where for fermionic atoms with half-integer nuclear spin, $2\mu$ is
always an odd integer, which implies that the sum in \eqref{eq:2tO_N+}
is an integer with the same parity (i.e.~even/odd) as the occupation
number $N$ (i.e.~the number of elements in $\N$).  At reduced times
$\tau_N=n\pi$ with integer $n$, therefore, if the occupation number
$N$ is even then all phases $2t\Omega_{\N,+}$ are integer multiples of
$2\pi$, which leads to a collapse of the excited-state populations as
$\left.\rho_N^{(\pm)}\p{t}\right|_{\tau_N=n\pi} = \rho_N^{(\pm)}\p{0}
= \rho_{N,0}$.  If the occupation $N$ is odd, meanwhile, then the
phases $2t\Omega_{\N,+}$ are all odd (even) integer multiples of $\pi$
for odd (even) $n$.  This alignment of phases implies a complete
population transfer to the excited state
$\rho_{N,-}\equiv\P_{N,-}/\tr\P_{N,-}$ for odd $n$, and a collapse
back to the orbital ground state $\rho_{N,0}$ for even $n$, precisely
as observed in figure \ref{fig:time}.

\section{Summary and outlook}
\label{sec:summary}

Current 3-D optical lattice experiments with fermionic AEAs are
capable of operating in the low-temperature, high-density,
strongly-interacting limit where inter-atomic interactions set the
dominant energy scale governing system dynamics.  For AEAs with total
nuclear spin $I$ and $N=2I+1$ nuclear spin states, these interactions
exhibit an exotic SU($N$) symmetry which is of great interest for
near-term quantum simulations of SU($N$) spin models and lattice field
theories.  Working in the deep-lattice limit and the experimental
regime of at most one atom occupying each nuclear spin state on any
lattice site, we have derived a low-energy effective theory of these
atoms.  Our theory exhibits emergent multi-body interactions that
inherit the SU($N$) symmetry of the bare two-body interactions.
Considering a restriction of our theory to the subspace of at most one
orbital excitation per lattice site, we found that the SU($N$)
symmetry of all $M$-body Hamiltonians allowed us to express them in a
simple form, and to fully characterize their eigenstates and spectra.
Capitalizing on the extreme precision of state-of-the-art clock
spectroscopy, we have tested spectral predictions of our theory
against direct experimental measurements of the many-body ${}^{87}$Sr
excitation spectrum.  This comparison shows good agreement between
theory and experiment, clearly demonstrating the need to consider
multi-body effects for understanding the low-energy physics of high
density AEA samples on a 3-D lattice.  Finally, we analyzed the
many-body orbital-state dynamics of multiply-occupied lattice sites
prepared in a nuclear spin mixture and interrogated via Rabi
spectroscopy.  This analysis is useful for future experimental probes
of many-body state structures, as well as for the preparation of
long-lived states with multi-partite entanglement (i.e.~$\ket{\N\pm}$)
which may be used as a resource to perform quantum information
processing tasks.

Despite the nominal success of our low-energy effective theory in
reproducing experimental observations, there remains room for
improvement in the form of controlled, systematic treatment of
higher-order and tunneling processes.  Nonetheless, our work makes a
major step towards the experimental investigation of multi-body
SU($N$) physics, providing the necessary framework for future studies
going beyond the deep-lattice limit to realize multi-body
super-exchange dynamics and orbital SU($N$) quantum magnetism with
AEAs.


\section*{Acknowledgements}

We acknowledge helpful discussions with R.~B.~Hutson, G.~E.~Marti,
S.~L.~Campbell, J.~Ye, P.~Julienne, J.~P.~D'Incao, C.~Kennedy, and
L.~Radzihovsky; and in particular close correspondence and technical
contributions from A.~Goban.  This work is supported by the Air Force
Office of Scientific Research (AFOSR) grant FA9550-18-1-0319; the
AFOSR Multidisciplinary University Research Initiative (MURI) grant;
the Defense Advanced Research Projects Agency (DARPA) and Army
Research Office (ARO) grant W911NF-16-1-0576; the National Science
Foundation (NSF) grant PHY-1820885; JILA-NSF grant PFC-173400; and the
National Institute of Standards and Technology (NIST).


\section*{Appendices}
\appendix
\setcounter{figure}{0}
\renewcommand\thefigure{\thesection.\arabic{figure}}

\section{Derivation of the effective Hamiltonian expansion}
\label{sec:eff_derivation}

Suppose we have a Hamiltonian $H_0$ on a Hilbert space
$\H=\G_0\oplus\E_0$ for a zero-energy manifold $\G_0$ decoupled from a
positive-energy manifold $\E_0$, and that we perturb $H_0$ by an
operator $V$ which weakly couples $\G_0$ and $\E_0$.  For all
$\ket{\psi_\g}\in\G_0$ and $\ket{\phi_\e},\ket{\chi_\e}\in\E_0$, we
have $\bk{\psi_\g|H_0|\psi_\g}=0$, $\bk{\phi_\e|H_0|\psi_\g}=0$, and
$\abs{\bk{\phi_\e|V|\psi_\g}}\ll\bk{\chi_\e|H_0|\chi_\e}$.  The net
Hamiltonian $H=H_0+V$ will naturally admit a decomposition of the
Hilbert space as $\H=\G\oplus\E$ for a subspace $\G$ which is spanned
by the low-energy eigenstates of $H$ and has the same dimension as
$\G_0$, i.e.~$\abs{\G}=\abs{\G_0}$.

We can perform a canonical transformation between $\G$ and $\G_0$
which yields an {\it effective Hamiltonian} $H_{\t{eff}}$ on $\G_0$
that reproduces the spectrum of $H$ on $\G$
\cite{bravyi2011schrieffer}.  Given an eigenbasis
$\set{\ket{\alpha_0}}$ for $H_0$ on $\G_0$ and $\set{\ket\alpha}$ for
$H$ on $\G$, this transformation is implemented by a unitary $U$ for
which $\ket{\alpha_0}=U\ket\alpha$ and $U\to\1$ as $\norm{V}\to0$.
The effective Hamiltonian is then simply
\begin{align}
  H_{\t{eff}} = U H U^\dag.
  \label{eq:H_eff}
\end{align}
The prescription in \eqref{eq:H_eff} for constructing an effective
Hamiltonian is commonly known as a Schieffer-Wolff transformation
\cite{schrieffer1966relation}.  Unitaries $U$ which follow this
prescription are not unique, and different choices of $U$ amount to
different realizations of the Schieffer-Wolff transformation.  In
ref.~\cite{bravyi2011schrieffer}, the authors construct the unique
operator $S$ which generates a {\it direct} or {\it minimal} rotation
$U_{\t{min}}=e^S$ between $\G$ and $\G_0$, and use this construction
to expand \eqref{eq:H_eff} as a perturbative series in $V$.  The
rotation $U_{\t{min}}$ is minimal in the sense that it minimizes the
distance of candiate unitaries $U$ from the identity $\1$ with respect
to the Euclidian operator norm\footnote{The Euclidean operator norm is
  also known as the $L_{2,2}$, Hilbert-Schmidt, or Frobenius norm.}
$\norm{X}_{\t{E}}\equiv\sqrt{\tr\p{X^\dag X}}$.  This rotation is
determined uniquely by enforcing
\begin{enumerate*}
\item that the generator $S$ is strictly block-off-diagonal with
  respect to $\G_0$ and $\E_0$,
\item that the norm $\norm{S}_{\t{E}}<\pi/2$, and
\item that the block-off-diagonal parts of \eqref{eq:H_eff} are zero.
\end{enumerate*}

To summarize the solution in ref.~\cite{bravyi2011schrieffer}, the
effective Hamiltonian $H_{\t{eff}}$ induced by a direct rotation can
be expanded as
\begin{align}
  H_{\t{eff}} = \sum_{p\ge0} H_{\t{eff}}^{(p)},
  \label{eq:H_eff_expansion}
\end{align}
where $H_{\t{eff}}^{(p)}$ is order $p$ in $V$.  Letting $\P_0$ denote
the projector onto $\G_0$, $\Q_0\equiv\1-\P_0$ denote the projector
onto $\E_0$, and $X$ denote any operator on $\H$, we define the
superoperators
\begin{align}
  \D X \equiv \P_0 X \P_0 + \Q_0 X \Q_0,
  &&
  \O X \equiv \P_0 X \Q_0 + \Q_0 X \P_0,
\end{align}
which select out the diagonal ($\D$) and off-diagonal ($\O$) parts of
$X$ with respect to $\G_0$ and $\E_0$, and
\begin{align}
  \L X \equiv \sum_{\alpha,\beta}
  \f{\op\alpha \O X \op\beta}{E_\alpha - E_\beta}
  &&
  \t{where}
  &&
  H_0 = \sum_\alpha E_\alpha \op\alpha.
\end{align}
The first few terms of the expansion in \eqref{eq:H_eff_expansion} are
then, as derived in ref.~\cite{bravyi2011schrieffer},
\begin{align}
  H_{\t{eff}}^{(0)} = \P_0 H_0 \P_0,
  &&
  H_{\t{eff}}^{(1)} = \P_0 V \P_0,
  \label{eq:H_eff_0_1}
\end{align}
\begin{align}
  H_{\t{eff}}^{(2)} = - \f12 \P_0 \sp{\O V,\L V} \P_0,
  &&
  H_{\t{eff}}^{(3)} = \f12 \P_0 \sp{\O V, \L \sp{\D V, \L V}} \P_0.
\end{align}
Exploiting the fact that in our case $\bk{\psi|H_0|\psi}=0$ for all
$\ket\psi\in\G_0$, we let $\B_0\p{\E_0}$ denote an eigenbasis of $H_0$
for $\E_0$ and define the operator
\begin{align}
  \I \equiv \sum_{\ket\alpha\in\B_0\p{\E_0}}\f{\op\alpha}{E_\alpha},
\end{align}
which sums over projections onto excited states with corresponding
energetic suppression factors.  We then expand
\begin{align}
  \L X = \O\p{\L X} = \Q_0 \L X \P_0 + \P_0 \L X \Q_0
  = \I X \P_0 - \P_0 X \I,
\end{align}
which simplifies the expression for $H_{\t{eff}}^{(2)}$ as
\begin{align}
  H_{\t{eff}}^{(2)}
  = - \f12 \P_0 \p{\sp{\O V, \I V \P_0}
    - \sp{\O V, \P_0 V \I}} \P_0
  = - \P_0 V \I V \P_0.
  \label{eq:H_eff_2}
\end{align}
Working toward a similar expansion for $H_{\t{eff}}^{(3)}$, we compute
\begin{align}
  \sp{\D V, \L V}
  = \sp{\D V, \I V \P_0} - \sp{\D V, \P_0 V \I}
  = \O\p{V \I V} - \I V \P_0 V \P_0 - \P_0 V \P_0 V \I,
\end{align}
and in turn
\begin{align}
  H_{\t{eff}}^{(3)}
  &= \f12 \P_0 \p{\sp{\O V, \I \sp{\D V, \L V} \P_0}
    - \sp{\O V, \P_0 \sp{\D V, \L V} \I}} \P_0 \\
  &= \f12 \P_0 \p{V \I \sp{\D V, \L V}
    + \sp{\D V, \L V} \I V} \P_0 \\
  &= \P_0 V \I V \I V \P_0
  - \f12 \P_0 V \I^2 V \P_0 V \P_0
  - \f12 \P_0 V \P_0 V \I^2 V \P_0 \\
  &= \P_0 V \I V \I V \P_0
  - \f12 \sp{\P_0 V \P_0, \P_0 V \I^2 V \P_0}_+,
  \label{eq:H_eff_3}
\end{align}
where $\sp{X,Y}_+\equiv XY+YX$.  The expressions in
\eqref{eq:H_eff_0_1}, \eqref{eq:H_eff_2}, and \eqref{eq:H_eff_3}
complete the derivation for our expansion of the effective interaction
Hamiltonian $H_{\t{int}}^{\t{eff}}$ in \eqref{eq:H_int_eff} through
third order.  In the case of ultracold atoms on a lattice, the
motional ground-state subspace $\G_0$ actually contains many internal
atomic states with different energies.  Nonetheless, the total Hilbert
space is completely separable into uncoupled subspaces associated with
each symmetrized many-body internal atomic state.  One can therefore
diagonalize the interaction Hamiltonian with respect to these internal
states and derive an effective theory within each of the corresponding
subspaces, in each case setting the appropriate ground-state energy to
zero.  This procedure is equivalent to simultaneously calculating the
effective Hamiltonian $H_{\t{int}}^{\t{eff}}$ for all internal states
via the prescriptions we have provided, but letting $E_\alpha$ denote
only the motional excitation energy of states
$\ket{\alpha_0}\in\B_0\p{\E_0}$.

\section{Diagram counting and symmetry factors}
\label{sec:diagrams}

The fact that we include factors of $1/2$ from the bare two-body
interaction Hamiltonian $H_{\t{int}}$ in the definition of diagrams
implies that $p$-vertex diagrams acquire a factor of $1/2^p$.  In
practice, however, these factors are exactly cancelled out by
corresponding symmetry factors in all $M$-body diagrams with $M>2$.
As an illustrative example, consider the second-order effective
Hamiltonian $H_{\t{int}}^{(2)}$ in \eqref{eq:H_int_1_2}, which
expanded in full reads
\begin{align}
  H_{\t{int}}^{(2)} = -\sum_{m+n>0} \P_0
  \p{\f12 K_{mn} G^{st}_{s't'}
    \c_{\rho s'}^\dag \c_{\sigma t'}^\dag \c_{n\sigma t} \c_{m\rho s}}
  \f1{E_{mn}}
  \p{\f12 K_{mn} G^{qr}_{q'r'}
    \c_{m\mu q'}^\dag \c_{n\nu r'}^\dag \c_{\nu r} \c_{\mu q}} \P_0.
  \label{eq:H_int_2_full}
\end{align}
The three-body terms in this Hamiltonian have
$\abs{\set{\mu,\nu,\rho,\sigma}}=3$ and only one virtually excited
atom.  The non-vanishing three-body terms must therefore either have
$\rho\in\set{\mu,\nu}$ and contain a factor of the form
$\c_X^\dag \c_{\sigma t'}^\dag \c_{\sigma t} \c_Y$, or have
$\sigma\in\set{\mu,\nu}$ with a factor of the form
$\c_{\rho s'}^\dag \c_X^\dag \c_Y \c_{\rho s}$, where the labels $X,Y$
both address whichever nuclear spin (i.e.~$\mu$ or $\nu$) was excited
in the corresponding term.  Diagrammatically, we have terms of the
form
\begin{align}
  \begin{tikzpicture}
    \begin{feynman}
      \vertex (v1);
      \vertex[above left = of v1] (f1) {$\mu q$};
      \vertex[below left = of v1] (f2) {$\nu r$};
      \vertex[right = of v1] (vm);
      \vertex[right = of vm] (v2);
      \vertex[above = of vm] (f3) {$Z$};
      \vertex[below = of vm] (f4) {$\sigma t$};
      \vertex[below right = of v2] (f5) {$\sigma t'$};
      \vertex[above right = of v2] (f6) {$X$};
      \diagram* {
        (f1) -- (v1) -- (f3),
        (f2) -- (v1) --  [scalar, edge label=$Y$] (v2),
        (f4) -- (v2) -- (f5),
        (v2) -- (f6), };
    \end{feynman}
  \end{tikzpicture}
  &&
  \t{and}
  &&
  \begin{tikzpicture}
    \begin{feynman}
      \vertex (v1);
      \vertex[above left = of v1] (f1) {$\mu q$};
      \vertex[below left = of v1] (f2) {$\nu r$};
      \vertex[right = of v1] (vm1);
      \vertex[right = of vm1] (vm2);
      \vertex[right = of vm2] (v2);
      \vertex[above = of vm1] (f3) {$Z$};
      \vertex[above = of vm2] (f4) {$\rho s$};
      \vertex[above right = of v2] (f5) {$\rho s'$};
      \vertex[below right = of v2] (f6) {$X$};
      \diagram* {
        (f1) -- (v1) -- (f3),
        (f2) -- (v1) --  [scalar, edge label=$Y$] (v2),
        (f4) -- (v2) -- (f5),
        (v2) -- (f6), };
    \end{feynman}
  \end{tikzpicture}.
  \label{eq:rho_sigma}
\end{align}
Observing that
$\c_{\rho s'}^\dag \c_X^\dag \c_Y \c_{\rho s} = \c_X^\dag \c_{\rho
  s'}^\dag \c_{\rho s} \c_Y$, however, it is clear that both of the
terms represented in \eqref{eq:rho_sigma} are equal up to the
re-indexing $\p{\sigma,t,t'}\leftrightarrow\p{\rho,s,s'}$.  There is
therefore a symmetry factor of $2$ associated with the second vertex
of the diagrams in \eqref{eq:rho_sigma}, which cancels out with the
explicit factor of $1/2$ at that vertex, i.e.~the first factor of
$1/2$ in \eqref{eq:H_int_2_full}.  A symmetry factor of essentially
identical origin appears at every vertex with an ``incoming'' virtual
state, as in e.g.~the second and third vertices of
\begin{align}
  \shrink{
    \begin{tikzpicture}
      \begin{feynman}
        \vertex (v1);
        \vertex[below right = of v1] (v2);
        \vertex[above right = of v2] (v3);
        \vertex[above left = of v1] (f1);
        \vertex[left = of v1] (f2);
        \vertex[below left = of v2] (f3);
        \vertex[above right = of v3] (f4);
        \vertex[right = of v3] (f5);
        \vertex[below right = of v2] (f6);
        \diagram* {
          (f1) -- (v1) --[scalar] (v3) -- (f4),
          (f2) -- (v1) --[scalar] (v2) --[scalar] (v3) -- (f5),
          (f3) -- (v2) -- (f6), };
      \end{feynman}
    \end{tikzpicture}}
  &&
  \t{and}
  &&
  \shrink{
    \begin{tikzpicture}
      \begin{feynman}
        \vertex (v1);
        \vertex[below right = of v1] (v2);
        \vertex[below right = of v2] (v3);
        \vertex[above left = of v1] (f1);
        \vertex[below left = of v1] (f2);
        \vertex[below left = of v2] (f3);
        \vertex[below left = of v3] (f4);
        \vertex[above right = of v1] (f5);
        \vertex[above right = of v2] (f6);
        \vertex[above right = of v3] (f7);
        \vertex[below right = of v3] (f8);
        \diagram* {
          (f1) -- (v1) -- (f5),
          (f2) -- (v1) -- (f5),
          (v1) --[scalar] (v2) -- (v3) -- (f8),
          (f3) -- (v2) -- (f6),
          (f4) -- (v3) -- (f7),
        };
      \end{feynman}
    \end{tikzpicture}},
  \label{eq:diagram_example_1}
\end{align}
or the last vertex of
\begin{align}
  \shrink{
    \begin{tikzpicture}
      \begin{feynman}
        \vertex (v1);
        \vertex[below right = 2.5em of v1] (v3);
        \vertex[below left = 1.7em of v3] (v2);
        \vertex[above left = of v1] (f1);
        \vertex[left = of v1] (f2);
        \vertex[left = of v2] (f3);
        \vertex[below left = of v2] (f4);
        \vertex[above right = of v1] (f5);
        \vertex[above right = of v3] (f6);
        \vertex[below right = of v3] (f7);
        \vertex[below right = of v2] (f8);
        \diagram* {
          (f1) -- (v1) -- (f5),
          (f2) -- (v1) --[scalar] (v3) -- (f6),
          (f3) -- (v2) -- (v3) -- (f7),
          (f4) -- (v2) -- (f8), };
      \end{feynman}
    \end{tikzpicture}}.
  \label{eq:diagram_example_2}
\end{align}
We can thus account for cancellations of $1/2$ at all vertices except
those which address two ``initial'' ground-state atoms, as in the
first vertex of the diagrams in \eqref{eq:rho_sigma} and
\eqref{eq:diagram_example_1}, or the first two vertices of the diagram
in \eqref{eq:diagram_example_2}.  For such vertices, there are two
possibilities: either
\begin{enumerate}
\item both edges leaving the vertex in question (i.e.~leaving to the
  right) terminate at different vertices, as in the examples above, or
  \label{enum:different_vertices}
\item both edges leaving the vertex in question terminate at the same
  vertex, as in for example the first vertex of
  \label{enum:same_vertex}
  \begin{align}
    \shrink{
      \begin{tikzpicture}
        \begin{feynman}
          \vertex (v1);
          \vertex[right = of v1] (v2);
          \vertex[below right = of v2] (v3);
          \vertex[above left = of v1] (f1);
          \vertex[below left = of v1] (f2);
          \vertex[below left = of v3] (f3);
          \vertex[above right = of v2] (f4);
          \vertex[above right = of v3] (f5);
          \vertex[below right = of v3] (f6);
          \diagram* {
            (f1) -- (v1) --[half left, scalar] (v2) -- (f4),
            (f2) -- (v1) --[half right, scalar] (v2)
            --[scalar] (v3) -- (f5),
            (f3) -- (v3) -- (f6), };
        \end{feynman}
      \end{tikzpicture}}.
    \label{eq:diagram_example_3}
  \end{align}
\end{enumerate}
In the former case, \ref{enum:different_vertices}, the vertex in
question has an associated symmetry factor of 2 to account for the
possibility of a nuclear spin exchange at that vertex.  Considering
again our example of the second-order effective Hamiltonian
$H_{\t{int}}^{(2)}$ in \eqref{eq:H_int_2_full}, the non-vanishing
three-body terms must either have $m=0$ and contain the factor
$\c_{\mu q'}^\dag \c_{n\nu r'}^\dag \c_{\nu r} \c_{\mu q}$, or have
$n=0$ with the factor
$\c_{m\mu q'}^\dag \c_{\nu r'}^\dag \c_{\nu r} \c_{\mu q}$, which
diagrammatically translates to
\begin{align}
  \begin{tikzpicture}
    \begin{feynman}
      \vertex (v1);
      \vertex[above left = of v1] (f1) {$\mu q$};
      \vertex[below left = of v1] (f2) {$\nu r$};
      \vertex[right = of v1] (vm);
      \vertex[right = of vm] (v2);
      \vertex[above = of vm] (f3) {$\mu q'$};
      \vertex[below = of vm] (f4) {$\sigma t$};
      \vertex[below right = of v2] (f5) {$\sigma t'$};
      \vertex[above right = of v2] (f6) {$\nu r''$};
      \diagram* {
        (f1) -- (v1) -- (f3),
        (f2) -- (v1) --  [scalar, edge label=$n\nu r'$] (v2),
        (f4) -- (v2) -- (f5),
        (v2) -- (f6), };
    \end{feynman}
  \end{tikzpicture}
  &&
  \t{or}
  &&
  \begin{tikzpicture}
    \begin{feynman}
      \vertex (v1);
      \vertex[above left = of v1] (f1) {$\mu q$};
      \vertex[below left = of v1] (f2) {$\nu r$};
      \vertex[right = of v1] (vm);
      \vertex[right = of vm] (v2);
      \vertex[above = of vm] (f3) {$\nu r'$};
      \vertex[below = of vm] (f4) {$\sigma t$};
      \vertex[below right = of v2] (f5) {$\sigma t'$};
      \vertex[above right = of v2] (f6) {$\mu q''$};
      \diagram* {
        (f1) -- (v1) -- (f3),
        (f2) -- (v1) --  [scalar, edge label=$m\mu q'$] (v2),
        (f4) -- (v2) -- (f5),
        (v2) -- (f6), };
    \end{feynman}
  \end{tikzpicture}.
  \label{eq:mu_nu}
\end{align}
These terms are equal up to the re-indexing
$\p{\nu,r,r',r'',n}\leftrightarrow\p{\mu,q,q',q'',m}$, which implies
that there is a symmetry factor of $2$ associated with the first
vertex of the diagrams in \eqref{eq:mu_nu}.  A symmetry factor of
identical origin is associated with the first vertex of the diagrams
in \eqref{eq:diagram_example_1}, and the first two vertices of the
diagram in \eqref{eq:diagram_example_2}.

The final case we must consider is \ref{enum:same_vertex}, which
occurs in the first vertex of \eqref{eq:diagram_example_3}.  In this
case, the symmetry factor of $2$ which appears in case
\ref{enum:different_vertices} to account for the possibility of a
nuclear spin exchange simply gets ``pushed forward'' to the vertex at
which the two nuclear spins in question part ways, e.g.~to account for
the two possibilities
\begin{align}
  \begin{tikzpicture}
    \begin{feynman}
      \vertex (v1);
      \vertex[right = of v1] (v2);
      \vertex[below right = of v2] (v3);
      \vertex[above left = of v1] (f1) {$\mu r$};
      \vertex[below left = of v1] (f2) {$\nu s$};
      \vertex[below left = of v3] (f3) {$\rho t$};
      \vertex[above right = of v2] (f4) {$\mu r''$};
      \vertex[above right = of v3] (f5) {$\nu s'''$};
      \vertex[below right = of v3] (f6) {$\rho t'$};
      \diagram* {
        (f1) -- (v1)
        --[half left, scalar, edge label=$\ell\mu r'$] (v2)
        -- (f4),
        (f2) -- (v1)
        --[half right, scalar, edge label'=$m\nu s'$] (v2)
        --[scalar, edge label=$n\nu s''$] (v3)
        -- (f5),
        (f3) -- (v3) -- (f6), };
    \end{feynman}
  \end{tikzpicture}
  &&
  \t{and}
  &&
  \begin{tikzpicture}
    \begin{feynman}
      \vertex (v1);
      \vertex[right = of v1] (v2);
      \vertex[below right = of v2] (v3);
      \vertex[above left = of v1] (f1) {$\mu r$};
      \vertex[below left = of v1] (f2) {$\nu s$};
      \vertex[below left = of v3] (f3) {$\rho t$};
      \vertex[above right = of v2] (f4) {$\nu s''$};
      \vertex[above right = of v3] (f5) {$\mu r'''$};
      \vertex[below right = of v3] (f6) {$\rho t'$};
      \diagram* {
        (f1) -- (v1)
        --[half left, scalar, edge label=$\ell\mu r'$] (v2)
        -- (f4),
        (f2) -- (v1)
        --[half right, scalar, edge label'=$m\nu s'$] (v2)
        --[scalar, edge label=$n\mu r''$] (v3)
        -- (f5),
        (f3) -- (v3) -- (f6), };
    \end{feynman}
  \end{tikzpicture},
  \label{eq:mu_nu_loop}
\end{align}
which are equal up to the re-indexing
$\p{\nu,s,s',s'',s''',m}\leftrightarrow\p{\mu,r,r',r'',r''',\ell}$.
The arguments for a symmetry factor in cases
\ref{enum:different_vertices} and \ref{enum:same_vertex} fail only if
the two nuclear spins in question take identical paths through the
internal vertices of a diagram, such that there is no meaningful sense
in which two diagrams can be said to differ by a nuclear spin
exchange, as in \eqref{eq:mu_nu} and \eqref{eq:mu_nu_loop}.  If two
atoms take identical paths through the internal vertices of a diagram,
however, then they have only participated in a two-body process, as in
\begin{align}
  \shrink{
    \begin{tikzpicture}
      \begin{feynman}
        \vertex (v);
        \vertex[above left = of v] (f1);
        \vertex[below left = of v] (f2);
        \vertex[above right = of v] (f3);
        \vertex[below right = of v] (f4);
        \diagram* {
          (f1) -- (v),
          (f2) -- (v),
          (v) -- (f3),
          (v) -- (f4) };
      \end{feynman}
    \end{tikzpicture}},
  &&
  \shrink{
    \begin{tikzpicture}
      \begin{feynman}
        \vertex (v1);
        \vertex[above left = of v1] (f1);
        \vertex[below left = of v1] (f2);
        \vertex[right = of v1] (v2);
        \vertex[above right = of v2] (f3);
        \vertex[below right = of v2] (f4);
        \diagram* {
          (f1) -- (v1),
          (f2) -- (v1),
          (v2) -- (f3),
          (v2) -- (f4),
          (v1) --[scalar, half left] (v2),
          (v1) --[scalar, half right] (v2) };
      \end{feynman}
    \end{tikzpicture}},
  &&
  \shrink{
    \begin{tikzpicture}
      \begin{feynman}
        \vertex (v1);
        \vertex[above left = of v1] (f1);
        \vertex[below left = of v1] (f2);
        \vertex[right = of v1] (v2);
        \vertex[right = of v2] (v3);
        \vertex[above right = of v3] (f3);
        \vertex[below right = of v3] (f4);
        \diagram* {
          (f1) -- (v1),
          (f2) -- (v1),
          (v3) -- (f3),
          (v3) -- (f4),
          (v1)
          --[scalar, half left] (v2)
          --[scalar, half left] (v3),
          (v1)
          --[scalar, half right] (v2)
          --[scalar, half right] (v3) };
      \end{feynman}
    \end{tikzpicture}},
  &&
  \cdots.
\end{align}
After summing over all free indices, therefore, all two-body diagrams
have a remaining factor of $1/2$ from the first vertex.  In all
connected $M$-body diagrams with $M>2$, meanwhile, every factor of
$1/2$ can be identified one-to-one with a corresponding symmetry
factor of $2$.

\section{Effective coupling constants in a lattice}
\label{sec:renormalization}

Due to our choice of renormalization scheme in section
\ref{sec:two_body}, the interaction energies prescribed by our
low-energy effective theory for multiply-occupied lattice sites are
not given directly by the coupling constants $G_X$ defined by the
free-space scattering lengths $a_X$ in \eqref{eq:couplings}.  Instead,
we must first compute effective coupling constants
$G_X^{\t{lattice}}\p{\U}$ in a lattice with depth $\U$, and in turn
use the effective coupling constants to compute interaction energies.
As the renormalization procedure $G_X\to G_X^{\t{lattice}}$ is
identical for all coupling constants, we henceforth drop the subscript
$X\in\set{\g\g,\e\g^-,\e\g^+,\e\e}$ on coupling constants $G_X$ in the
remainder of this Appendix.  To further simplify notation, we will
also neglect the explicit dependence of parameters on the lattice
depth $\U$, which we generally keep fixed.

Proper calculations of the interaction energy of two ultracold
fermions in an optical lattice were performed in
refs.~\cite{wall2013strongly} and \cite{buchler2010microscopic} using
a two-channel model of a Feshbach resonance, yielding prescriptions
for computing effective coupling constants in a lattice from
free-space interaction parameters.  These calculations, however, are
both analytically and numerically involved.  We therefore instead opt
to use a modified version of the considerably simpler single-channel
calculation in ref.~\cite{busch1998two} of the interaction energy of
two ultracold atoms in a harmonic trap.  Our approach is equivalent to
the calculation of Hubbard parameters performed in
ref.~\cite{dickerscheid2005feshbach}, and has been demonstrated to
reproduce correct results in the limit of a deep lattice (compared to
the lattice photon recoil energy) and small positive scattering
lengths (compared to the effective harmonic oscillator length)
\cite{buchler2010microscopic}.

The exact result in Eq. 16 of ref.~\cite{busch1998two} for the
interaction energy of two ultracold atoms in a harmonic oscillator
with angular trap frequency $\omega$ can be written in the form
\begin{align}
  \p{G_{\t{free}} K_{\t{HO}}/\omega}^{-1}
  = \f{\sqrt{\pi}~ \Gamma\p{-G_{\t{HO}}K_{\t{HO}}/2\omega}}
  {\Gamma\p{-G_{\t{HO}}K_{\t{HO}}/2\omega-1/2}},
  &&
  K_{\t{HO}} \equiv \int \d^3x~ \abs{\phi_0^{\t{HO}}}^4,
  \label{eq:harmonic_gamma}
\end{align}
where $G_{\t{HO}}$ is an effective coupling constant in the harmonic
trap, $\phi_0^{\t{HO}}\p{x}$ is the corresponding non-interacting
ground-state wavefunction, and $\Gamma$ is the gamma function.  The
expression in \eqref{eq:harmonic_gamma} can be solved numerically as
is, or expanded about $G_{\t{HO}}K_{\t{HO}}/\omega=0$ to get
\begin{align}
  G_{\t{free}}^{-1} = G_{\t{HO}}^{-1}
  \sum_{n=0}^\infty c_n \p{G_{\t{HO}}K_{\t{HO}}/\omega}^n,
  \label{eq:harmonic_series}
\end{align}
where the first few coefficients are
\begin{align}
  c_0 = 1,
  &&
  c_1 = 1 - \ln 2,
  &&
  c_2 = -\f{\pi^2}{24} - \ln 2 + \f12 \p{\ln 2}^2.
\end{align}
The series in \eqref{eq:harmonic_series} can in turn be inverted to
solve for $G_{\t{HO}}$ with an expansion of the form
\begin{align}
  G_{\t{HO}} = G_{\t{free}}
  \sum_{n=0}^\infty \tilde c_n \p{G_{\t{free}}K_{\t{HO}}/\omega}^n,
  \label{eq:inverted_series}
\end{align}
where if we truncate the series in \eqref{eq:harmonic_series} at
$n=2$, the first few coefficients of \eqref{eq:inverted_series} are
\begin{align}
  \tilde c_0 = 1,
  &&
  \tilde c_1 = 1 - \ln 2,
  &&
  \tilde c_2 = -\f{\pi^2}{24} - \ln 2 + \f12 \p{\ln 2}^2
  + \p{1 - \ln 2}^2.
  \label{eq:inverted_coefficients}
\end{align}
The coefficients $\tilde c_n$ thus found are consistent with the
coefficients $c_2^{(n+1)}$ reported in table 1 of
ref.~\cite{johnson2012effective}, in which the authors compute the
first few terms of the two-body Hamiltonian $H_2$ directly as
expressed in \eqref{eq:H_2_expansion} by using a renormalization
scheme which subtracts off divergences term by term.

All of the above results are exact for two atoms in a harmonic
oscillator interacting via $s$-wave scattering.  In order to adapt
these results for a lattice, we expand the lattice potential about a
lattice site centered at $x=(0,0,0)$ as
\begin{align}
  \U\sin^2\p{k_{\t{L}} \cdot x}
  \approx \U \sp{\p{k_{\t{L}}^\x k_{\t{L}}^\y k_{\t{L}}^\z}^{1/3} x}^2
  = \f12 \mA \omega_{\t{eff}}^2 x^2,
  &&
  \omega_{\t{eff}} \equiv \sqrt{2~ \U k_{\t{L}}^2 / \mA},
\end{align}
where $k_{\t{L}}=\p{k_{\t{L}}^\x,k_{\t{L}}^\y,k_{\t{L}}^\z}$ is the
lattice wavenumber, $\mA$ is the atomic mass, and $\omega_{\t{eff}}$
is an effective angular harmonic trap frequency.  We then use
$\omega_{\t{eff}}$ in place of $\omega$ in \eqref{eq:harmonic_gamma},
and use an overlap integral $K$ computed with the ground-state
wavefunctions $\phi_0$ in a lattice rather than those in a harmonic
oscillator.  We retrieve free-space $s$-wave scattering lengths
$a_{\t{free}}$ for ${}^{87}$Sr from ref.~\cite{zhang2014spectroscopic}
to determine the free-space coupling constants
$G_{\t{free}} \equiv \p{4\pi/\mA} a_{\t{free}}$.  This procedure
yields an effective coupling constant $G_{\t{lattice}}$ given by
\begin{align}
  \p{G_{\t{free}} K/\omega_{\t{eff}}}^{-1}
  = \f{\sqrt{\pi}~ \Gamma\p{-G_{\t{lattice}}K/2\omega_{\t{eff}}}}
  {\Gamma\p{-G_{\t{lattice}}K/2\omega_{\t{eff}}-1/2}},
\end{align}
with a solution
\begin{align}
  G_{\t{lattice}} = G_{\t{free}}
  \sum_{n=0}^\infty \tilde c_n \p{G_{\t{free}}K/\omega_{\t{eff}}}^n,
\end{align}
where the first few coefficients are provided in
\eqref{eq:inverted_coefficients}.

\section{Momentum-dependent $s$-wave interactions}
\label{sec:momentum_dependence}

In addition to the renormalization of coupling constants discussed in
Appendix \ref{sec:renormalization}, computing two-body interaction
energies $E_{NX}^{(2)}$ at third order in the low-energy effective
theory requires accounting for the contribution of momentum-dependent
$s$-wave interactions.  At next-to-leading order in the relative
momentum $k$ between two atoms, the effective momentum-dependent
scattering length $a_{\t{eff}}$ is given in terms of the zero-momentum
scattering length $a$ by \cite{giorgini2008theory, blume2002fermi,
  flambaum1999analytical}
\begin{align}
  \f1{a_{\t{eff}}}
  = \f1a - \f12 r_{\t{eff}} k^2
  = \f1a\p{1 - \f12 r_{\t{eff}} a k^2},
\end{align}
which for $r_{\t{eff}} a k^2 \ll 1$, implies that
\begin{align}
  a_{\t{eff}}
  \approx a\p{1 + \f12 r_{\t{eff}} a k^2}
  = a + \f12 r_{\t{eff}} a^2 k^2.
  \label{eq:a_eff}
\end{align}
Here $r_{\t{eff}}$ is an effective range of $\O\p{k^2}$ interactions,
determined in atomic units by the scattering length $a$ and van der
Waals $C_6$ coefficient by \cite{flambaum1999analytical}
\begin{align}
  r_{\t{eff}} = \f13 \xi^{-2} \chi \p{1 - 2\chi + 2\chi^2} a,
  &&
  \t{where}
  &&
  \xi \equiv \f{\Gamma\p{3/4}}{\Gamma\p{1/4}},
  &&
  \chi \equiv \sqrt{2}~ \xi~ \f{\p{\mA C_6}^{1/4}}{a},
  \label{eq:effective_range}
\end{align}
and $\Gamma$ is the gamma function.  As $\chi\sim1$ for ${}^{87}$Sr,
the momentum-dependent correction to the effective scattering length
$a_{\t{eff}}$ is $\O\p{a^3}$ without an additional separation of
scales (i.e.~which could have occurred if we had $\chi\ll1$ or
$\chi\gg1$).  The momentum-independent $\O\p{k^0}$ contribution to
$a_{\t{eff}}$ in \eqref{eq:a_eff} gives rise to the bare two-body
interactions in \eqref{eq:H_int} by use of an unregularized contact
(i.e.~$\delta$-function) potential, while the momentum-dependent
$\O\p{k^2}$ term gives rise to the interaction Hamiltonian
\cite{blume2002fermi, johnson2012effective}
\begin{align}
  H_{\t{int}}'
  \equiv \f12 \sum {G'}^{qr}_{st} \int \d^3 x~d^3y~\delta\p{z}
  \sp{\hat\psi_{\mu q}^\dag\p{x} \hat\psi_{\nu r}^\dag\p{y}}
  \hat k_z^2\sp{\hat\psi_{\nu t}\p{y} \hat\psi_{\mu s}\p{x}},
  \label{eq:H_int_primed}
\end{align}
where
\begin{align}
  z \equiv x - y,
  &&
  \hat k_z^2
  \equiv -\f12 \p{\vec\nabla_z^2 + \lvec\nabla_z^2},
\end{align}
and the primed couplings ${G'}^{qr}_{st}$ are defined similarly to
unprimed couplings $G^{qr}_{st}$ in \eqref{eq:couplings} and
\eqref{eq:coupling_tensor}, but with scattering lengths
$a\to r_{\t{eff}}a^2/2$ and the effective range $r_{\t{eff}}$ defined
by \eqref{eq:effective_range} for each scattering length with an
appropriate $C_6$ coefficient.  We retrieve $C_6$ coefficients for
${}^{87}$Sr from the supplementary material of
ref.~\cite{zhang2014spectroscopic}.  The squared relative momentum
operator $k_z^2$ is represented by symmetrized left- and right-acting
derivative operators in order to preserve manifest Hermiticity of
$H_{\t{int}}'$.  At third order in the low-energy effective theory
developed in section \ref{sec:low_energy}, the bare momentum-dependent
interactions in \eqref{eq:H_2} yield only the effective two-body
Hamiltonian
\begin{align}
  H_2' \equiv \f12 K' \sum {G'}^{qr}_{st}
  \c_{\mu s}^\dag \c_{\nu t}^\dag \c_{\nu r} \c_{\mu q},
  \label{eq:H_2_prime}
\end{align}
where, letting $\Re\sp{X}$ denote the real part of $X$,
\begin{align}
  K' \equiv \f12 \int \d^3x~
  \Re\sp{\p{\phi_0^*}^2
    \p{\vec\nabla\phi_0\cdot\vec\nabla\phi_0
      - \phi_0\vec\nabla^2\phi_0}}.
\end{align}

\section{Bounds on theoretical uncertainties from inter-site effects}
\label{sec:error}

In our overview of the relevant one- and two-particle physics of
ultracold atoms on a lattice (section \ref{sec:overview}), we made two
approximations which introduce error into the low-energy effective
theory.  Both approximations concern the on-site locality of the
single- and two-body Hamiltonians: we assumed that
\begin{enumerate*}
\item tunneling between lattice sites and
\item inter-site interactions are negligible.
\end{enumerate*}
These approximations are justified for single-particle motional ground
states of atoms in a deep lattice, but generally break down when
considering virtual states occupying highly excited motional levels,
whose spatial wavefunctions can span multiple lattice sites.
Nonetheless, we can place upper bounds on the magnitude of inter-site
corrections to the effective on-site interaction Hamiltonians by
treating tunneling and inter-site interactions of virtual excited
states perturbatively and assuming no energetic penalty for
nearest-neighbor hopping.  These bounds can be used to diagnose the
breakdown of the on-site effective theory, and signal when a more
careful consideration of inter-site effects is necessary to make
precise predictions about many-body spectra and dynamics.

If we still assume negligible overlap between single-particle
ground-state wavefunctions in different lattice sites but consider
nearest-neighbor wavefunction overlaps of states with motional
excitations, our one-body and bare two-body Hamiltonians become
\begin{align}
  H_0 = \sum E_n \c_{in\mu s}^\dag \c_{in\mu s}
  - \sum_{\substack{\bk{i,j}\\m,n>0}}
  \p{t_{mn} \c_{jn\mu s}^\dag \c_{im\mu s} + \t{h.c.}},
  \label{eq:H_0_neighbor}
\end{align}
and
\begin{align}
  H_{\t{int}} &= \f12 \sum K^{k\ell}_{mn} G^{qr}_{st}
  \c_{im\mu s}^\dag \c_{in\nu t}^\dag \c_{i\ell\nu r} \c_{ik\mu q}
  + \f12 \sum_{\substack{\bk{i,j}\\n>0}} G^{\mu q;\nu r}_{\rho s;\sigma t}
  \p{\K_n \c_{j,0,\rho s}^\dag \c_{j,0,\sigma t}^\dag
    \c_{j,0,\nu r} \c_{in\mu q} + \t{h.c.}} \nonumber \\
  &\quad + \f12 \sum_{\substack{\bk{i,j}\\m,n>0}}
  G^{\mu q;\nu r}_{\rho s;\sigma t}
  \p{\K_{mn} \c_{j,0,\rho s}^\dag \c_{j,0,\sigma t}^\dag
    \c_{in\nu r} \c_{im\mu q} +
    \tilde\K_{mn} \c_{in\rho s}^\dag \c_{j,0,\sigma t}^\dag
    \c_{j,0,\nu r} \c_{im\mu q} + \t{h.c.}},
  \label{eq:H_int_neighbor}
\end{align}
where $t_{mn}$ is a tunneling rate; $\K_n,\K_{mn},\tilde\K_{mn}$ are
inter-site spatial overlap integrals; $\t{h.c.}$ denotes a Hermitian
conjugate, i.e.~$\p{X+\t{h.c.}}\equiv \p{X+X^\dag}$; and $\bk{i,j}$
denotes the set of all lattice sites $i$ together with their adjacent
sites $j$.  Note that we have neglected terms in
\eqref{eq:H_int_neighbor} which involve more than two field operators
addressing states with motional excitations, as these terms will not
appear in the leading-order corrections to the effective on-site
interaction Hamiltonians.  We also still neglect terms which involve
products of atomic wavefunctions for motional ground states in
different lattice sites.

Diagrammatically representing matrix elements of $H_0$ and
$H_{\t{int}}$ which are off-diagonal in lattice site by a dot
(i.e.~$\bullet$) and marking lines which represent field operators
addressing neighboring lattice sites by a cross (i.e.~$+$ or $\times$,
depending on the line orientation), the dominant terms in the
effective theory which we previously neglected by assuming on-site
locality are
\begin{align}
  \begin{tikzpicture}
    \begin{feynman}
      \vertex (v1);
      \vertex[above left = of v1] (f1);
      \vertex[below left = of v1] (f2);
      \vertex[right = of v1] (v2);
      \vertex[above right = of v2] (f3);
      \vertex[below right = of v2] (f4);
      \diagram* {
        (f1) -- (v1),
        (f2) -- (v1),
        (v1) --[scalar, half left] (v2),
        (v1) --[scalar, half right] (v2),
        (v2) --[insertion = 0.5] (f3),
        (v2) --[insertion = 0.5] (f4) };
      \draw[fill=black] (v2) circle(0.8mm);
    \end{feynman}
  \end{tikzpicture}
  &\sim \gamma_2^{(2)} G^2,
  & \gamma_2^{(2)} &\equiv \sum_{n+m>0} \f{K_{mn}\K_{mn}}{E_{mn}},
  \label{eq:g_2_2} \\[1em]
  \begin{tikzpicture}
    \begin{feynman}
      \vertex (v1);
      \vertex[above left = of v1] (f1);
      \vertex[below left = of v1] (f2);
      \vertex[right = 4em of v1] (v2);
      \vertex[above right = of v1] (f3);
      \vertex[below left = of v2] (f4);
      \vertex[below right = of v2] (f5);
      \vertex[above right = of v2] (f6);
      \diagram* {
        (f1) -- (v1) -- (f3),
        (f2) -- (v1) --[scalar] (v2),
        (f4) --[insertion = 0.5] (v2) --[insertion = 0.5] (f5),
        (v2) --[insertion = 0.5] (f6), };
      \draw[fill=black] (v2) circle(0.8mm);
    \end{feynman}
  \end{tikzpicture}
  &\sim \gamma_{3,1}^{(2)} G^2,
  & \gamma_{3,1}^{(2)} &\equiv \sum_{n>0} \f{K_n\K_n}{E_n},
  \label{eq:g_3_1_2} \\[1em]
  \begin{tikzpicture}
    \begin{feynman}
      \vertex (v1);
      \vertex[above left = of v1] (f1);
      \vertex[below left = of v1] (f2);
      \vertex[right = 2em of v1] (v2);
      \vertex[right = 2em of v2] (v3);
      \vertex[above right = of v1] (f3);
      \vertex[below left = of v3] (f4);
      \vertex[below right = of v3] (f5);
      \vertex[above right = of v3] (f6);
      \diagram* {
        (f1) -- (v1) -- (f3),
        (f2) -- (v1) --[scalar] (v2) --[scalar, insertion = 0.5] (v3),
        (f4) --[insertion = 0.5] (v3) --[insertion = 0.5] (f5),
        (v3) --[insertion = 0.5] (f6), };
      \draw[fill=black] (v2) circle(0.8mm);
    \end{feynman}
  \end{tikzpicture}
  &\sim \gamma_{3,2}^{(2)} G^2,
  & \gamma_{3,2}^{(2)} &\equiv \sum_{n,m>0} \f{K_mt_{mn}K_n}{E_mE_n},
  \label{eq:g_3_2_2}
\end{align}
where we have identified, up to an assignment of coupling constants
$G$, the magnitude of all nonzero matrix elements of the diagrams with
respect to an eigenbasis of the on-site single-particle Hamiltonian
$H_0$ in \eqref{eq:H_0}.

The terms in \eqref{eq:g_2_2}-\eqref{eq:g_3_2_2} can be used to
estimate an upper bound on the magnitude of dominant corrections to
the spectrum of the low-energy theory from off-diagonal (i.e.~in
lattice site) matrix elements of the Hamiltonians in
\eqref{eq:H_0_neighbor} and \eqref{eq:H_int_neighbor}.  Conservatively
assuming no energetic penalty and no Pauli blocking for any inter-site
process, the dominant correction $\delta E_N$ to the interaction
energy of a lattice site with $N$ atoms and $b$ neighboring sites
(e.g.~$b=6$ in a primitive cubic lattice) is roughly bounded as
\begin{align}
  \abs{\delta E_N} \lesssim b {N\choose 2}
  \max\set{\abs{\gamma_2^{(2)}},
    \p{N-1} \abs{\gamma_{3,1}^{(2)} + \gamma_{3,2}^{(2)}}} G^2,
  \label{eq:delta_E_N}
\end{align}
where the factor of $b$ accounts for the multiplicity of neighboring
sites; the factor of ${N\choose2}$ accounts for the number of on-site
pairs of atoms which are addressed by the diagrams in
\eqref{eq:g_2_2}-\eqref{eq:g_3_2_2}; and the factor of $N-1$ on
$\gamma_{3,X}^{(2)}$ accounts for the number of atoms in a neighboring
site which are addressed by the corresponding processes.  These
factors count the number of matrix elements in the Hamiltonian with
magnitude $\sim\gamma_X^{(2)}G^2$.  The maximization in
\eqref{eq:delta_E_N} is performed because the relevant two- and
three-body processes are mutually exclusive, requiring a different
number of atoms on neighboring lattice sites.  For a conservative
bound of $\abs{\delta E_N}$, the coupling factor $G^2$ in
\eqref{eq:delta_E_N} can simply be maximized over its allowed values
for a given state of atoms on a lattice site, e.g.~$G_\g^2$ for a
state with no orbital excitations, or $\max\set{G_\g^2,G_+^2,G_-^2}$
for a state with one net orbital excitation (in both cases, assuming
no orbital excitations in neighboring sites).  In the latter case, the
bound in \eqref{eq:delta_E_N} can also be reduced by observing that to
conserve energy, it must be the excited atom which moves to a
neighboring site, which reduces the factor of ${N\choose2}$ in down to
$N-1$.  We emphasize that the bound in \eqref{eq:delta_E_N} is by no
means an exact measure of error, and is merely intended to provide a
conservative range of energies and corresponding time scales for which
inter-site effects could become relevant despite negligible
single-particle ground-state tunneling rates.

\section{Perturbative parameters for the effective theory}
\label{sec:pert_params}

\begin{figure}
  \centering
  \includegraphics{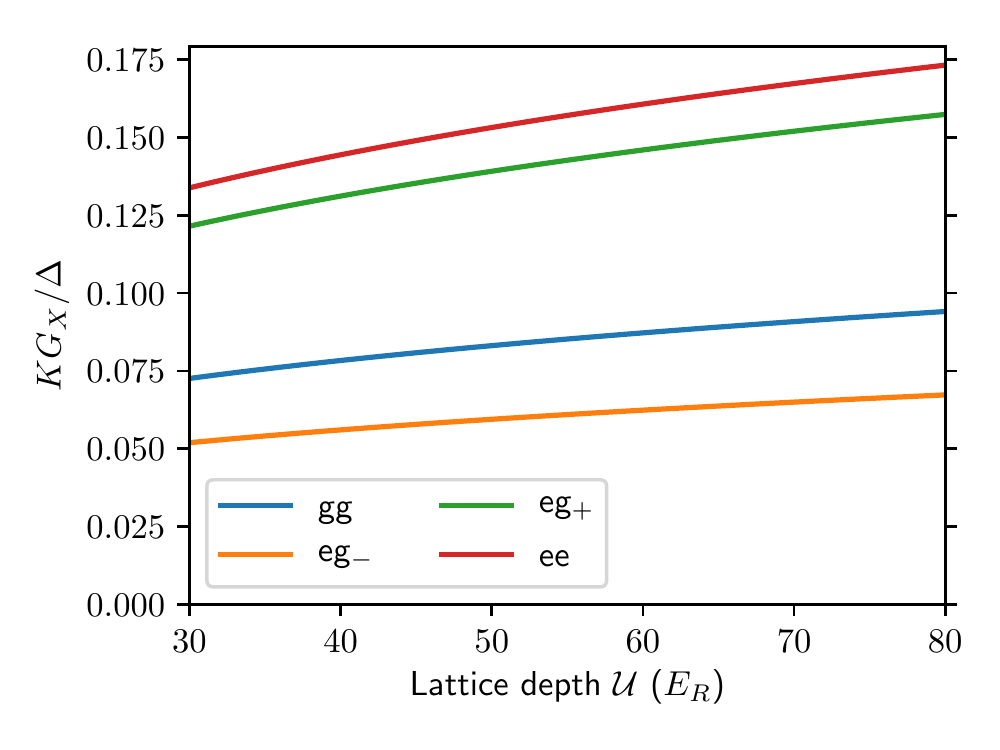}
  \caption{\footnotesize Dependence of the perturbative parameters
    $KG_X/\Delta$ on the lattice depth $\U$.}
  \label{fig:pert_params}
\end{figure}

The perturbative effective theory developed in Section
\ref{sec:low_energy} is organized in powers of the coupling constants
$G_X$.  The formally small, dimensionless quantities for this
perturbation theory are the two-body interaction energies $K G_X$
divided by the spectral gap $\Delta$ of the non-interacting
Hamiltonian $H_0$.  Here $K$ is a ground-state two-body overlap
integral and $G_X$ is a coupling constant.  The quantities $K$, $G_X$,
and $\Delta$ all depend on the lattice depth $\U$.  Figure
\ref{fig:pert_params} shows these parameters for the case of
${}^{87}$Sr atoms with $X\in\set{\g\g, \e\g_-, \e\g_+, \e\e}$ at
lattice depths $\U\in\sp{30,80}E_{\t{R}}$, where
$E_{\t{R}}\approx3.5\times2\pi~\t{kHz}$ is the lattice photon recoil
energy of the atoms.  The fact that these perturbative parameters grow
with increasing lattice depth $\U$ is a consequence of the fact that
the overlap integral $K$ grows faster with $\U$ than the spectral gap
$\Delta$.  In the case of a harmonic trap with angular frequency
$\omega$, for example, by dimensional analysis these parameters would
be
\begin{align}
  \f{K_{\t{HO}} G_X}{\omega}
  = \f{G_X}{\omega} \int \d^3 x~\abs{\phi_{\t{HO}}}^4
  = \f{G_X}{\omega} \sp{\int \d x~
    \abs{\p{\f{\mA\omega}{\pi}}^{1/4} e^{-\mA\omega x^2/2}}^4}^3
  \propto \sqrt{\omega},
\end{align}
where we assumed that the coupling constants $G_X$ vary weakly with
$\omega$.  While this result may seem to suggest that the low-energy
effective theory should become better at smaller lattice depths,
smaller lattice depths also result in increased theoretical
uncertainties from the growing relevance of the inter-site effects
discussed in Appendix \ref{sec:error}.

\section{Low-excitation $M$-body Hamiltonian coefficients}
\label{sec:U_X}

When restricted to the subspace of at most one orbital excitation per
lattice site, the $M$-body Hamiltonians of the low-energy effective
theory developed in Section \ref{sec:low_energy} can be written in the
form
\begin{align}
  H_M = \sum_{\abs{\set{\mu_j}}=M}
  \p{U_{M,\g} \n_{\mu_1,\g} \n_{\mu_2,\g}
    + U_{M,+} \n_{\mu_1,\e} \n_{\mu_2,\g}
    + U_{M,-} \c_{\mu_1,\g}^\dag \c_{\mu_2,\e}^\dag
    \c_{\mu_2,\g} \c_{\mu_1,\e}}
  \prod_{\alpha=3}^M \n_{\mu_\alpha,\g},
\end{align}
where the coefficients can be expanded as
$U_{MX} = \sum_p U_{MX}^{(p)}$ with terms $U_{MX}^{(p)}$ at order $p$
in the coupling constants $G_Y$.  The terms $U_{MX}^{(p)}$ can be
determined from the $M$-body $p$-order Hamiltonians $H_M^{(p)}$
derived in section \ref{sec:low_energy}, i.e.~in \eqref{eq:H_2},
\eqref{eq:H_3_2}, \eqref{eq:H_3_3}, and \eqref{eq:H_4_3}.  For the
effective 2-, 3-, and 4-body Hamiltonians through third order in the
coupling constants, the coefficients are
\begin{align}
  U_{2,\g} = \f12 \alpha_2^{(1)} G_\g,
  &&
  U_{2,+} = \alpha_2^{(1)} G_+,
  &&
  U_{2,-} = \alpha_2^{(1)} G_-,
\end{align}
\begin{align}
  U_{3,\g}^{(2)} = - \alpha_3^{(2)} G_\g^2,
  &&
  U_{3,+}^{(2)} = - \alpha_3^{(2)} G_+ \p{G_+ + 2 G_\g},
\end{align}
\begin{align}
  U_{3,-}^{(2)} = - \alpha_3^{(2)} G_- \p{2 G_+ + G_- + 2 G_\g},
\end{align}
\begin{align}
  U_{3,\g}^{(3)}
  = \p{\alpha_{3,1}^{(3)} - \alpha_5^{(3)}} 2 G_\g^3
  + \p{2\alpha_{3,2}^{(3)} - \alpha_{4,3}^{(3)} - \alpha_5^{(3)}} G_\g^3,
\end{align}
\begin{align}
  U_{3,+}^{(3)}
  &= \p{\alpha_{3,1}^{(3)} - \alpha_5^{(3)}} \p{G_+^3 + 4 G_+^2 G_\g
    + G_+ G_-^2 + G_+ G_\g^2 + G_-^3 + G_-^2 G_\g} \nonumber \\
  &\quad + \p{2\alpha_{3,2}^{(3)} - \alpha_{4,3}^{(3)} - \alpha_5^{(3)}}
  \p{G_+^3 + G_+^2 G_\g + G_+ G_-^2 + G_+ G_\g^2 + G_-^2 G_\g},
\end{align}
\begin{align}
  U_{3,-}^{(3)}
  &= \p{\alpha_{3,1}^{(3)} - \alpha_5^{(3)}} G_- \p{3 G_+^2
    + 2 G_+ G_- + 8 G_+ G_\g + 3 G_- G_\g + G_\g^2} \nonumber \\
  &\quad + \p{2\alpha_{3,2}^{(3)} - \alpha_{4,3}^{(3)} - \alpha_5^{(3)}}
  G_- \p{3 G_+^2 + 2 G_+ G_- + 2 G_+ G_\g + G_-^2 + G_\g^2},
\end{align}
\begin{align}
  U_{4,\g}^{(3)}
  = \p{2\alpha_{4,1}^{(3)} - \alpha_5^{(3)}} G_\g^3
  + \p{\alpha_{4,2}^{(3)} - \alpha_5^{(3)}} 2 G_\g^3,
\end{align}
\begin{align}
  U_{4,+}^{(3)}
  &= \p{2\alpha_{4,1}^{(3)} - \alpha_5^{(3)}}
  2 G_+ G_\g \p{G_+ + G_\g}
  + \p{\alpha_{4,2}^{(3)} - \alpha_5^{(3)}}
  G_+ \p{G_+^2 + 2 G_+ G_\g + 5 G_\g^2},
\end{align}
\begin{align}
  U_{4,-}^{(3)}
  &= \p{2\alpha_{4,1}^{(3)} - \alpha_5^{(3)}}
  2 G_- G_\g \p{2 G_+ + G_- + G_\g} \nonumber \\
  &\quad + \p{\alpha_{4,2}^{(3)} - \alpha_5^{(3)}}
  G_- \p{3 G_+^2 + 3 G_+ G_- + 4 G_+ G_\g + G_-^2
    + 2 G_- G_\g + 5 G_\g^2},
\end{align}
In terms of the spatial overlap integrals defined in \eqref{eq:K_klmn}
and \eqref{eq:K}, the prefactors $\alpha_X^{(p)}$ on the coefficients
$U_X^{(p)}$ are
\begin{align}
  \alpha_2^{(1)} \equiv K,
  &&
  \alpha_3^{(2)} \equiv \sum_{n>0} \f{K_n^2}{E_n},
  &&
  \alpha_5^{(3)}
  \equiv  K \sum_{n>0} \f{K_n^2}{E_n^2},
\end{align}
\begin{align}
  \alpha_{3,1}^{(3)} \equiv \sum_{\substack{\ell+m>0\\\ell+n>0}}
  \f{K_{\ell m} K^m_n K_{\ell n}}{E_{\ell m} E_{\ell n}},
  &&
  \alpha_{3,2}^{(3)}
  \equiv \sum_{\substack{\ell+m>0\\n>0}}
  \f{K_{\ell m} K_n}{E_{\ell m} E_n}
  \p{K^{\ell m}_n - \f{K_{\ell m} K_n}{K}},
\end{align}
\begin{align}
  \alpha_{4,1}^{(3)}
  \equiv \sum_{\substack{m\ge0\\n>0}} \f{K_{mn} K_m K_n}{E_{mn} E_n},
  &&
  \alpha_{4,2}^{(3)}
  \equiv \sum_{m,n>0} \f{K_m K^m_n K_n}{E_m E_n},
  &&
  \alpha_{4,3}^{(3)}
  \equiv K \sum_{m+n>0} \f{K_{mn}^2}{E_{mn}^2}.
\end{align}

\bibliography{\jobname}

\end{document}